\documentclass[conference]{IEEEtran}
\IEEEoverridecommandlockouts

\usepackage{cite}
\usepackage{algorithmic}
\usepackage{textcomp}
\usepackage{booktabs}
\usepackage{multirow}
\usepackage{graphicx}
\usepackage{xspace}
\usepackage{pifont}
\usepackage{amsmath,amssymb,amsfonts}
\usepackage{listings}
\usepackage{xcolor}
\usepackage{tikz}
\usepackage{siunitx}
\usepackage{enumitem}

\usepackage{subcaption}
\usepackage{balance}
\usepackage{url}

\renewcommand{\paragraph}{\vspace{2pt}\noindent \textbf}

\def\UrlBreaks{\do\/\do-}

\newcommand{\ecall}{{$\tt{ECALL}$}\xspace}

\newcommand{\ocall}{{$\tt{OCALL}$}\xspace}
\newcommand{\ecalls}{{$\tt{ECALL}$s}\xspace}
\newcommand{\ocalls}{{$\tt{OCALL}$s}\xspace}
\newcommand{\libc}{{$\tt{libc}$}\xspace}
\newcommand{\glibc}{{$\tt{glibc}$}\xspace}
\newcommand{\musl}{{$\tt{musl}$}\xspace}

\newcommand{\gsgx} {{Graphene-SGX}\xspace}
\newcommand{\tool} {{\sc Ratel}\xspace}
\newcommand{\ratel} {{\sc Ratel}\xspace}
\newcommand{\codename} {{\sc Ratel}\xspace}
\newcommand{\dr} {{DynamoRIO}\xspace}

\newcommand*\ver[1]{\hbox to1em{\hss\rotatebox[origin=br]{90}{#1}}}
\providecommand{\loc}{LoC\xspace}

\definecolor{lightgray}{rgb}{.9,.9,.9}
\definecolor{darkgray}{rgb}{.4,.4,.4}
\definecolor{purple}{rgb}{0.65, 0.12, 0.82}

\lstdefinelanguage{JavaScript}{
  keywords={typeof, static, new, begin, end, struct, char, void, unsigned, long, const, privilege_enclave, int, true, false, catch, function, return, null, catch, switch, var, if, in, while, do, else, case, break},
  keywordstyle=\color{blue},
  ndkeywords={class, export, boolean, throw, implements, import, this},
  ndkeywordstyle=\color{darkgray}\bfseries,
  identifierstyle=\color{black},
  sensitive=false,
  comment=[l]{//},
  morecomment=[s]{/*}{*/},
  commentstyle=\color{purple}\ttfamily\bfseries,
  stringstyle=\color{red}\ttfamily,
  morestring=[b]',
  morestring=[b]"
}

\lstdefinestyle{JavaScript}{
    language={JavaScript}, 
    moredelim=**[is][\btHL]{`}{`},
    moredelim=**[is][{\btHL[fill=green!30,]}]{@}{@},
}

\lstdefinestyle{nonumbers}
{numbers=none}

\def\BibTeX{{\rm B\kern-.05em{\sc i\kern-.025em b}\kern-.08em
    T\kern-.1667em\lower.7ex\hbox{E}\kern-.125emX}}
\begin{document}

\title{Binary Compatibility For SGX Enclaves}
\author{\IEEEauthorblockN{Shweta Shinde$^{\dagger}$}
\IEEEauthorblockA{UC Berkeley}
~\thanks{$^{\dagger}$Part of the research was done while at National University of Singapore.}
\and
\IEEEauthorblockN{Jinhua Cui}
\IEEEauthorblockA{National University of Singapore \& \\ National University of Defense Technology}
\and
\IEEEauthorblockN{Satyaki Sen \qquad Pinghai Yuan \qquad Prateek Saxena}
\IEEEauthorblockA{National University of Singapore}
}

\maketitle

\begin{abstract}
Enclaves, such as those enabled by Intel SGX, offer a powerful
hardware isolation primitive for application partitioning. To become
universally usable on future commodity OSes, enclave designs should
offer compatibility with existing software.  In this paper, we draw
attention to $5$ design decisions in SGX that create incompatibility
with existing software. These represent concrete starting points, we
hope, for improvements in future TEEs. Further, while many prior works
have offered partial forms of compatibility, we present the first
attempt to offer {\em binary compatibility} with existing software on
SGX.  We present \tool, a system that enables a dynamic binary
translation engine inside SGX enclaves on Linux. Through the lens of
\tool, we expose the fundamental trade-offs between performance and
complete mediation on the OS-enclave interface, which are rooted in
the aforementioned $5$ SGX design restrictions. We report on an
extensive evaluation of \tool on over $200$ programs, including
micro-benchmarks and real applications such as Linux utilities.
\end{abstract}
 \section{Introduction}
\label{sec:intro}

Commercial processors today have native support for trusted execution
environments (TEEs) to run user-level applications in isolation from
other software on the system. A prime example of such a TEE is Intel
Software Guard eXtensions (SGX)~\cite{sgx}. The hardware-isolated environment
created by SGX, commonly referred to as an {\em enclave}, runs a
user-level application without trusting privileged software. 

Enclaved TEEs offer a powerful new foundation for compartmentalization
on commodity OSes. Enclaves remove privileged software layers (OS or
hypervisor) from the trusted code base of isolated components.
Therefore, they crucially differ from existing isolation primitives
like processes, virtual machines, and containers.
Enclaves offer the intriguing possibility of becoming ubiquitously
used abstractions in the future, much like processes, but this demands
a scale of usage not originally envisioned with SGX. Enclaves
would need to be compatible with a large fraction of the existing
software and OS abstractions. Arguably, compatibility is the most
important challenge facing future enclaved TEEs. One would only be
concerned with additional enclave security threats if they could run
the desired application in the first place.

Right from the inception, compatibility with existing x86\_64 binaries
has been a recognized issue with SGX~\cite{sgx-explained}. 
Prior works have proposed a number of different ways of achieving
partial compatibility---offering specific programming languages for
authoring enclave code~\cite{open-enclave, asylo}, keeping
compatibility with container interfaces~\cite{scone, ryoan}, or
conformance to specific versions of library interfaces provided by
library OSes~\cite{graphene-sgx, haven, sgx-lkl, panoply, occlum}.

While these approaches are promising and steadily maturing, none of
them offer {\em binary compatibility} with existing software. In
existing approaches, applications are expected to be relinked against
specific versions of libraries (e.g., $\tt{musl}$, $\tt{libc}$,
$\tt{glibc}$), ported to a customized OS, or containerized. Such
modifications require significant changes to the complex build systems
in place, often demanding developer involvement and even access to
source code. 
More importantly, most prior works have enabled sufficient SGX
compatibility to handle specific applications~\cite{tor-sgx, sgx-ml,
vc3-followup-ccs}, standard libraries, or language
runtimes~\cite{civet, sgx-lang-interpreter, acctee, go-tee, trustjs}
that these platforms choose. Trade-offs arising between compatibility,
security, and performance are often resolved in the favor of
performance in prior designs. Thus, the complete picture of these
fundamental trade-offs has never been presented.

The purpose of this paper is two-fold. First, we believe that future
designs of enclave TEEs would benefit from understanding which design
choices made in SGX create binary incompatibility. In particular, we
pinpoint at 5 specific SGX restrictions and explain how they create a
sweeping incompatibility with existing OS abstractions of
multi-threading, memory mapping, synchronization,
signal-handling, shared memory, and others. These challenges affect
other compatibility approaches too. However, our emphasis on full
binary compatibility brings them out more comprehensively.

Second, we study the feasibility of a new approach that can offer
binary compatibility for unmodified applications in SGX enclaves. Our
approach enables interposition of enclave applications via {\em
dynamic binary translation} (DBT). DBT is a mature technique for
cross-platform binary compatibility available even before SGX~\cite{dynamorio}. It
works by instrumenting machine instructions on-the-fly to provide a
layer of transparency to the underlying system. This is sometimes
referred to as application-level virtualization.

We report on our experience of running a widely used DBT framework
called \dr inside SGX enclaves~\cite{bruening-thesis}. The resulting
system called \codename enables DBT on Intel SGX
enclaves for unmodified x86\_64 Linux binaries. \codename aims to
ensure that it adheres to enclave threat model: it does not trust the
OS in its design. The challenges of enabling a full-fledged DBT engine
is instructive in exposing a set fundamental trade-offs on SGX---one
has to choose ``$2$ out of $3$'' between security, binary
compatibility, and performance. \tool chooses to resolve these
in favor of security and binary compatibility.

The main challenge addressed by \tool is that it offers {\em secure
and complete mediation} on all data and control flow between the OS
and the application. Our main finding is that the problem of complete
mediation suffers from the ``last mile'' phenomenon: We pay modestly
in performance to get to partial mediation, as seen in many prior
works, but significantly for the full binary compatibility on SGX.

As a standalone tool, \tool offers its own conceptual utility. It
offers complete mediation on all application-OS interactions, which is
useful for security interposition. This side steps challenges of
static source-code based solutions, which expect changes ahead of
time, and can work for dynamically generated or self-modifying code.
Further, DBT provides the facility of instruction-level
instrumentation. This can be useful in many ways: inlining security
monitors, sandboxing, fine-grained resource accounting, debugging, or
deployment of third-party patches in response to newly discovered
flaws (though these are not our focus)~\cite{dbo, program-shepherding,
dyncfi, dr-memory, dr-shadow-stack}.

\paragraph{Contributions \& Results.} \tool is the first system that
targets binary compatibility for SGX, to the best our knowledge. Our
proposed design enables an industrial-strength dynamic binary
translation engine inside SGX enclaves. We evaluate compatibility
offered by \codename extensively. We successfully run a total of
$203$ unique unmodified binaries across $5$ benchmark suites ($58$
binaries), $4$ real-world application use-cases ($12$ binaries), and
$133$ Linux utilities. These encompass various work-load profiles
including CPU-intensive (SPEC 2006), I/O system call intensive (FSCQ,
IOZone), system stress-testing (HBenchOS), multi-threading support
(Parsec-SPLASH2), a machine learning library (Torch), and real-world 
applications demonstrated in prior works on SGX. \codename
offers compatibility but does not force applications to use any
specific libraries or higher-level interfaces.

Our work pin-points $5$ specific design choices in SGX that are
responsible for incompatibility. We believe these create challenges
for prior approaches providing partial compatibility as well. We hope
future enclave TEE designs pay attention to addressing these $5$
points from the outset. 
 \section{Why is Binary Compatibility Challenging?}
\label{sec:prob}

Intel SGX allows execution of code inside a hardware-isolated
environment called an enclave~\cite{sgx-explained}.\footnote{Unless
stated otherwise, we use the term Intel SGX v1 to refer to the
hardware as well as the trusted platform software (PSW) and the
trusted software development kit (SDK), as shown in
Figure~\ref{fig:dr}.} 
SGX enforces confidentiality and integrity of enclave-bound code and
data.
All enclave memory is private and only accessible when executing in
enclave-mode. Data exchanged with the external world (e.g., the host
application or OS) must reside in public memory which is not
protected.
At runtime, one can only synchronously enter an enclave via \ecalls
and exit an enclave via \ocalls. Any illegal instructions or
exceptions in the enclave create asynchronous entry-exit points. SGX
restricts enclave entry and exits to pre-specified points in the
program. 
If the enclave execution is interrupted asynchronously, SGX saves the
enclave context and resumes  it at the same program point at a later
time. 
Our challenge is to interpose securely and completely on all the control
and data transfers between the enclave and the OS.
 
\subsection{Restrictions Imposed by SGX Design}
\label{sec:restrictions}

Intel SGX protects the enclave by enforcing strict isolation at
several points of interactions between the OS and the user code. We
outline $5$ SGX design restrictions.

\begin{enumerate}[label=\textbf{R\arabic*.}]
\item {\bf Spatial memory partitioning.}
SGX enforces {\em spatial} memory partitioning. 
It reserves a region that is private to the enclave and
the rest of the virtual memory is public. Memory can either be public
or private, not both. 
 
\item {\bf Static memory partitioning.}
The enclave has to specify the spatial partitioning {\em statically}.
The size, type (e.g., code, data, stack, heap), and permissions for
its private memory have to be specified before creation and these
attributes cannot be changed at runtime. 

\item {\bf Non-shareable private memory.}
An enclave cannot share its private memory with other enclaves on the
same machine.

\item {\bf 1-to-1 private virtual memory mappings.}
Private memory spans over a contiguous virtual address range, whose
start address is decided by the OS. Private virtual address has
one-to-one mapping with a physical address. 

\item {\bf Fixed entry points.} 
Enclaves can resume execution only from its last point and context of
exit. Any other entry points/contexts have to be statically
pre-specified as valid ahead of time.

\end{enumerate}

\begin{table}[tbp]
\centering
\resizebox{0.40\textwidth}{!}{
\begin{tabular}{@{}ll@{}}
\toprule
\begin{tabular}[c]{@{}l@{}}OS\\ abstraction\end{tabular} & 
\begin{tabular}[c]{@{}l@{}}Restrictions\\ Affecting Abstraction\end{tabular} \\ 
\midrule
 System call arguments        & R1             \\
 Dynamic Loaded / Gen. code   & R2             \\
 Thread Support               & R5, R2         \\
 Signal Handling              & R1, R5         \\
 Thread Synchronization       & R3, R1         \\
 File / Memory Mapping        & R1, R2, R3, R4 \\
 IPC / Shared Memory          & R3, R4         \\
\bottomrule
\end{tabular}
}
\caption{Ramifications of SGX design restrictions on common OS abstractions.}
\label{tab:restrictions}
\end{table} 
\subsection {Ramifications}
\label{sec:ramifications}

Next, we explain the impact of these design restrictions on various OS
and application functionality (see Table~\ref{tab:restrictions}).

\paragraph{R1.} 
Since SGX spatially partitions the enclave memory, any data which is
exchanged with the OS requires copying between private and public
memory. In normal applications, an OS assumes that it can access all
the memory of a user process, but this is no longer true for enclaves.
Any arguments that point to enclave private memory are not accessible
to the OS or the host process. The enclave has to explicitly manage a
public and a private copy of the data to make it accessible externally
and to shield it from unwanted modification when necessary. We refer
to this as a {\em two-copy mechanism}. 
Thus, $R1$ breaks functionality (e.g., system calls, signal handling,
futex), introduces non-transparency (e.g., explicitly synchronizing
both copies), and introduces security gaps (e.g., TOCTOU
attacks~\cite{toctou, iago}).

\paragraph{R2.} 
Applications often require changes to the size or permissions of
enclave memory. For example, memory permissions change after dynamic
loading of libraries (e.g., $\tt{dlopen}$) or files (e.g.,
$\tt{mmap}$), executing dynamically generated code, creating
read-only zero-ed data segments (e.g., {.bss}), and for software-based
isolation of security-sensitive data. The restriction R2 is
incompatible with such functionality. To work with this restriction,
applications require careful semantic changes: either weaken the
protection (e.g., read-and-execute instead of read-or-execute), use a
two-copy design, or rely on some additional form of isolation (e.g.,
using segmentation or software instrumentation).

\paragraph{R3.} 
SGX has no mechanism to allow two enclaves to share parts of their
private memory directly. This restriction is incompatible with the
synchronization primitives like locks and shared memory when there is
no trusted OS synchronization service. Keeping two copies of locks
breaks the semantics and create a chicken-and-egg issue: how to
synchronize two copies of a shared lock without another trusted
synchronization primitive.

\paragraph{R4.} 
When applications demand new virtual address mappings (e.g., malloc)
the OS adds these mappings. The application can ask the OS to map the
same physical page at several different offsets---either with same or
different permissions. For example, the same file is mapped as
read-only at two places in the program. Since, SGX doesn't allow the
same physical address to be mapped to multiple virtual addresses, any
such mappings generate a general protection fault in SGX. 

\paragraph{R5.} 
SGX starts or resumes enclave execution only from controlled entry
points i.e., the virtual address and the execution context. However,
there are several unexpected entry points to an application when we
run them unmodified in an enclave (e.g., exception handlers, library
functions, illegal instructions). Statically determining all potential
program points for re-entry is challenging. 
Further, when the enclave resumes execution, it expects the same
program context.
This does not adhere with typical program functionality. Normally, if
the program wants to execute custom error handling code, say after a
divide-by-zero ($\tt{SIGFPE}$) or illegal instruction ($\tt{SIGILL}$),
it can resume execution at a handler function in the binary with
appropriate execution context setup by the OS. 
On the contrary, SGX will resume enclave execution at the same
instruction and same context (not the OS setup context for exception
handling), thus re-triggering the exception.

Intel is shipping SGX v2, wherein an enclave can make dynamic changes
to private page permissions, type, and size. We discuss their
specifics in Section~\ref{sec:related}. Note that, SGX v2 only
addresses R2 partially, while all the other restrictions still hold
true. Thus, for the rest of the paper, we describe our design based
on SGX v1. \section{Overview}
\label{sec:overview}

Before we present our design, we emphasize our key
empirical takeaway that led to it: {\em Working with restrictions $R1-R5$, we are faced with a ``choose
2-out-of-3'' trilemma between security, performance, and binary
compatibility}. We explain these trade-offs in Section~\ref{sec:tradeoffs}.
Our design picks security and compatibility over performance, wherever
necessary. In this design principle, it fundamentally departs from prior work.

\begin{figure}[tbp]
\centering
\includegraphics[width=0.45\textwidth]{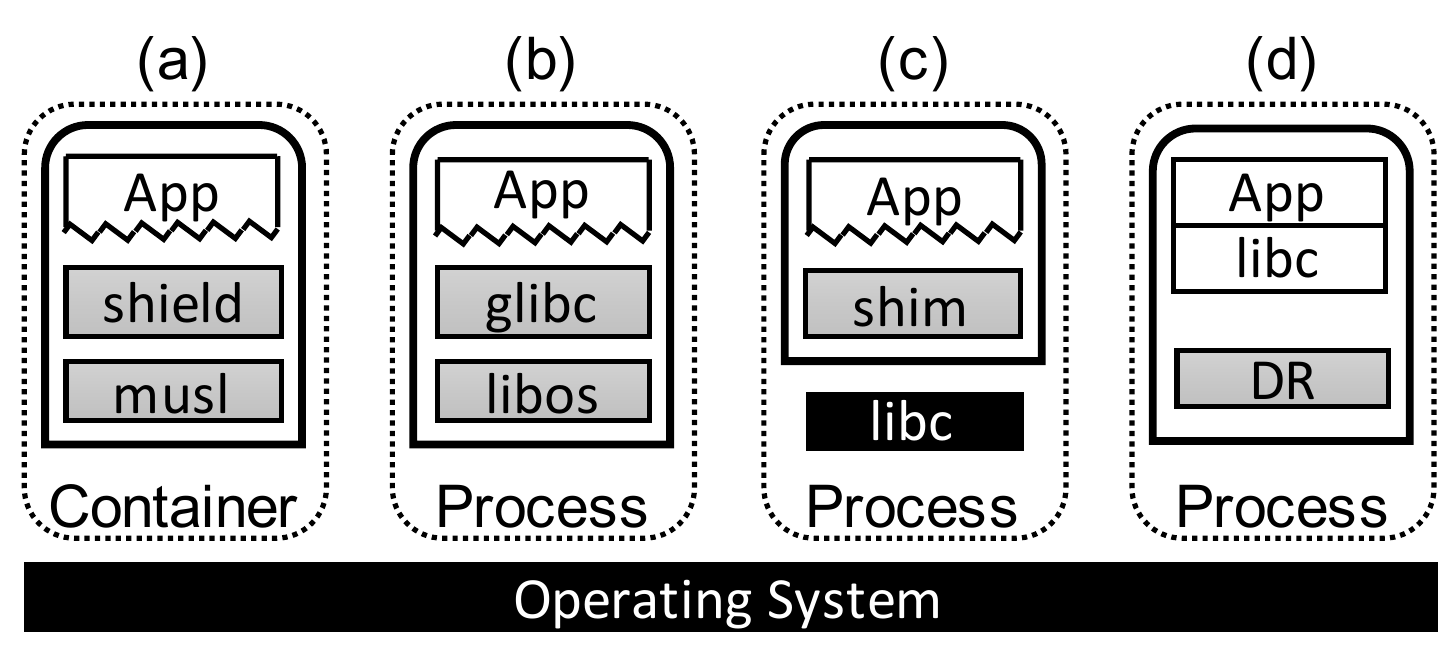}
\caption{Different abstraction layers for compatibility. Black shaded
regions are untrusted, gray shaded regions are modifications or
additions, thick solid lines are enclave boundaries, clear boxes are
unmodified components, zig-zag lines show break in compatibility.
(a) Container abstraction with \musl ~\libc interface (Scone~\cite{scone}).
(b) Library OS with \glibc interface (\gsgx~\cite{graphene-sgx}).
(c) Process abstraction with POSIX interface (Panoply~\cite{panoply}).
(d) Dynamic Binary Translation with \dr in \tool (This work).
}
\label{fig:api-levels}
\end{figure}

Several different approaches to enable applications in SGX enclaves
have been proposed. In nearly all prior works, performance
consideration dominate design decisions. A prominent way to side-step
the performance costs of ensuring compatibility and complete mediation
is to ask the application to use a prescribed program-level interface
or API. The choice of interfaces vary. They include specific
programming languages~\cite{fortanix-rust-sgx,
  baidu-rust-sgx,go-tee,civet}, application
frameworks~\cite{tensorscone}, container interfaces~\cite{scone}, and
particular implementation of standard \libc interfaces.
Figure~\ref{fig:api-levels} shows the prescribed interfaces in three
approaches, including library OSes and container engines, and where
they intercept the application to maintain compatibility. Given that
binary compatibility is not the objective of prior works, they handle
subsets of $R1-R5$. One drawback of these approaches is that if an
application does not originally use the prescribed API, the
application needs to be rewritten, recompiled from source, or relinked
against specific libraries.

Our work poses the following question: Can full {\em binary
  compatibility} be achieved on SGX, and if so, with what trade-offs
in security or performance? Application binaries are originally
created with the intention of running on a particular OS and we aim to
retain compatibility with the OS system call interface
(e.g., Linux). In concept, applications are free to use any library,
direct assembly code, and runtime that uses the Linux system call
interface. The central challenge we face is to enable \textbf{\em
  secure and complete mediation of all data and control flow} between
the application and the OS. We do this by enabling a widely used DBT
engine inside enclaves.

\subsection{Background: Dynamic Binary Translation}
\label{sec:dbt}

Dynamic binary translation is a well-known approach to achieving full
binary compatibility. It was designed to secure and complete
mediation: the ability to intercept each instruction in the program
before it executes. DBT works by first loading the binary code that is
about to be executed into its own custom execution engine. It then
updates the code in-situ, if required, and then dispatches it for
execution. To contrast it with the approach of changing \libc, DBT
intercepts the application right at the point at which it interacts
with the OS (see Figure~\ref{fig:api-levels}).

\begin{figure}
\center
\includegraphics[width=0.33\textwidth]{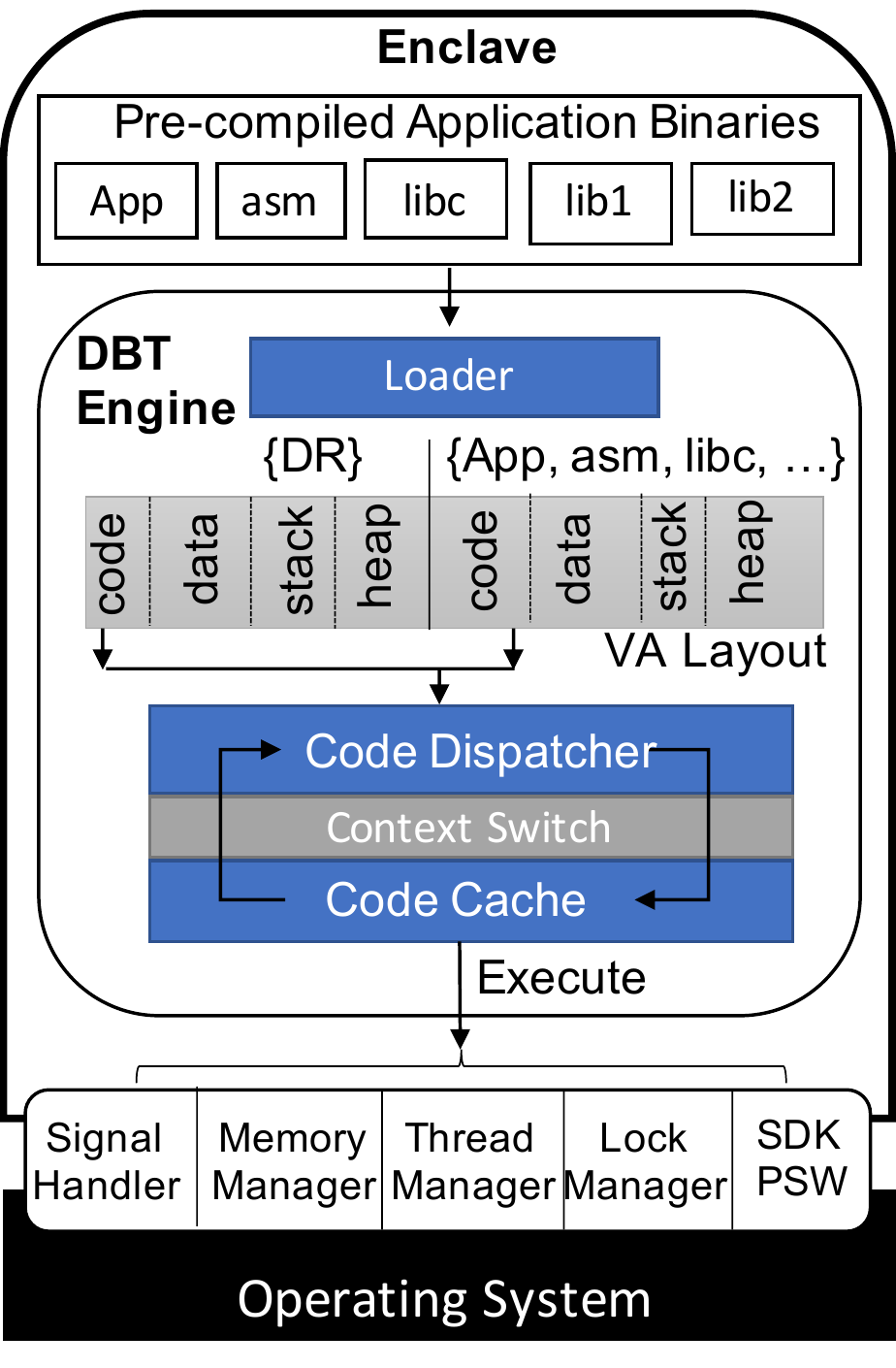}
\caption{\codename overview.
}
\label{fig:dr}
\end{figure}

In this paper, we choose \dr as our DBT engine, since it is
open-source and widely used in industry~\cite{dynamorio}.\footnote{The
other option is Intel Pin~\cite{pintool}, but it is not open-source.} 
\dr is itself an example of just-in-time compilation engine which
dynamically generates code.  At a high level, \dr first loads itself
and then loads the application code in a separate part of the virtual
address, as show in Figure~\ref{fig:dr}. Similarly, it sets up two
different contexts, one for itself and one for the application.
\dr can update the code on-the-fly before putting it in the code-cache
by re-writing instructions (e.g., convert syscall instruction to a
library function call). Such rewriting ensures that \dr engine takes
control before each block of code executes, enabling the ability to
interpose on every instruction. Instrumented code blocks are placed in
a region of memory called a code cache. When the code cache executes,
\dr regains control as the instrumentation logic desires. It does
post-execution updates to itself for book-keeping or to the program's
state. Additionally, \dr hooks on all events intended for the process
(e.g., signals). The application itself is prevented from accessing \dr
memory via address-space isolation
techniques~\cite{program-shepherding}. Thus, it acts as an arbiter
between the application's binary code and the external environment
(e.g., OS, filesystem) with secure and complete mediation.

\codename retains the entire low-level instruction translation and
introspection machinery of \dr, including the code cache and its
performance optimizations. This enables reusing well-established
techniques for application transparency, instrumentation, and
performance enhancements. We eliminate the support for auxiliary
plugins to reduce TCB.

\subsection{\codename Approach}
\label{sec:tradeoffs}

Our design must provide compatibility for both the \dr DBT engine as
well as any application binary code that runs translated. We provide a
high-level overview of \tool and then explain the key trade-offs we
face when forced with compliance to SGX restrictions $R1-R5$. 

\paragraph{High-level Overview.}
We modify \dr to adhere to SGX virtual memory limitations
(R1-R4) by setting up our custom layout. Specifically, we analyze \dr
code to identify its entire virtual address layout. 
This allows us to load \codename and start its
execution without violating the memory semantics of SGX.
We register a fixed entry point in \codename when entering or resuming
the enclave. This entry point acts as a unified trampoline, such that
upon entry, \codename decides where to redirect the control flow,
depending on the previously saved context.
In \dr code, we manually replace instructions that are illegal in
SGX with an external call that executes outside the enclave. Thus,
\codename execution itself is guaranteed to never violate R5.

The same challenges show up when \codename starts loading and running
the translated application binary. However, we have the advantage of
dynamically rewriting the application logic to adapt it to R1-R5.
We statically initialize the virtual memory size of the application
to the maximum allowed by SGX; the type and permissions of memory is
set to the specified type in the binary. 
We add a memory management unit to \dr to keep track of and
transparently update the applications layout.
At runtime, when the application makes direct changes to its own
virtual memory layout via system calls, \codename dynamically adapts
it to SGX (e.g., by making two copies or relocating the virtual
mappings). \codename intercepts all application interactions with the
OS. It modifies application parameters, OS return values, OS events
for monitoring indirect changes to the memory (e.g., thread creation).
In the other direction, \codename also intercepts OS events on the
behalf of the application. Upon re-entry, if the event has to be
delivered to the application, it sets/restores the appropriate
execution context and resumes execution via the trampoline.
Lastly, before executing any application logic, \codename scans the
code cache for any instructions (e.g., $\tt{syscall}$, $\tt{cpuid}$)
that may potentially be deemed as illegal in SGX and replaces it with
an external call. Thus, \codename remedies the application on-the-fly
to adhere to R1-R5.

\paragraph{Key Design Trade-offs.}
\tool aims for {\em secure and complete} mediation on all data
and control flow between the application and OS, through the use of
DBT. This makes \tool useful for a wide variety of reasons:
in-lining security monitoring, software sandboxing, and even profiling
and debugging. We do not assume that the application is written to
help \tool by adhering to restrictions beyond those specified by a
normal OS, nor do we trust the OS. SGX restrictions $R1-R5$  give rise
to trade-offs between security, compatibility, and performance. We
point out that these are somewhat fundamental and apply to \tool and
other compatibility efforts equally. However, \tool chooses security
and compatibility over performance, whenever conflicts between the
three arise.

Whenever the application reads from or writes outside the enclave, the
data needs to be placed in public memory due to $R1$. Computing on data
in public memory, which is exposed to the OS, is insecure. Therefore,
if the application wishes to securely compute on the data, a copy must
necessarily be maintained in a separate private memory space, as $R2$
forbids making changing data permissions. This leads to a ``two copy''
mechanism, instances of which repeat throughout the design. The
two-copy mechanism, however, incurs both space and computational
performance overheads, as data has to be relocated at runtime.
Further, certain data structure semantics which require a single
memory copy (e.g. \texttt{futex}es) become impossible to keep
compatibility with (see Section~\ref{sec:synch}).

$R3$ creates an ``all or none'' trust model between enclaves. Either
memory is shared with all entities (including the OS) or kept private
to one enclave. $R4$ restricts sharing memory within an enclave
further. These restrictions are in conflict with semantics of shared
memory and synchronization primitives. To implement such abstractions
securely, the design must rely on a trusted software manager which
necessarily resides in an enclave, since the OS is
untrusted. Applications can then have compatibility with lock and
shared memory abstractions and securely, but at the cost of
performance: Access to shared services turn into procedure calls to
the trusted manager.

Restriction $R5$ requires that whenever the enclave
resumes control after an exit, the enclave state (or context) should
be the same as right before exit. This implies that the security
monitor (e.g., the DBT engine) must take control before all exit points
and after resumption, to save-restore contexts---otherwise, the
mediation can be incomplete, creating security holes and
incompatibility. Without guarantees of complete mediation, the OS can
return control into the enclave, bypassing a security checks that the
DBT engine implements. The price for complete mediation on binaries is
performance: the DBT engine must intercept all entry/exit points and
simulate additional context switches in software. Prior approaches,
such as library-OSes, sacrifice complete mediation (security) for
better performance, by asking applications to link against specific
library interfaces which tunnel control via certain points. But, this
does not enforce complete mediation as applications can make direct OS
interactions or override entry handlers, intentionally or by oversight. 
There are several further security considerations that arise in the
details of the above design decisions. These include (a) avoiding
na{\"i}ve designs that have TOCTOU attacks; (b) saving and restoring
the execution context from private memory; (c) maintaining
\codename-specific metadata in private memory to ensure integrity of
memory mappings that change at runtime; and (d) explicitly zeroing out
memory content and pointers after use. We explain them inline in
Section~\ref{sec:design}.

\subsection{Scope}
\label{sec:scope}

Many challenges are common between the design of \tool presented here
and other systems. These include encryption/decryption of external
file or I/O content~\cite{scone, ryoan, graphene, occlum},
sanitization of OS inputs to prevent Iago attacks~\cite{iago, besfs,
coins, tale-two-worlds}, defenses against known side-channel
attacks~\cite{pigeonhole,cosmix,tsgx,drsgx,sgx-cache-transmem},
additional attestation or integrity of dynamically loaded/generated
code~\cite{acctee,go-tee,sgx-lang-interpreter, trustjs}, and so on.
These are important but largely orthogonal to the focus of this work.
These can be implemented on top of \tool in the future.

Our focus is to expose the compatibility challenges that SGX creates
with rich process-level abstractions. These require careful design to
eliminate additional security threats. One limitation of \tool is that
 the present implementation of \tool has support for majority but not
all of the Linux system calls. The most notable of the unsupported
system calls  is $\tt{fork}$ which is used for multi-processing. We
believe that the basic design of \tool can be extended to support
$\tt{fork}$ with the two-copy mechanism, similar to prior
work~\cite{panoply}. Our experience suggests that adding other system
calls is a tedious but conceptually straight-forward effort in \tool.
We expect to expand the syscall coverage over time, possibly with the
help of automated tools.

  \section{\codename Design}
\label{sec:design}

Our main challenge is to execute the system functionality securely
while faithfully preserving its semantics for compatibility. We
explain how \codename design achieves this for various sub-systems that
are typically expected by applications. 

\subsection{System Calls \& Unanticipated Entry-Exits}
\label{sec:syscall}

SGX does not allow enclaves to execute several instructions such as
$\tt{syscall}$, $\tt{cpuid}$, $\tt{rdtsc}$. If the enclave executes
them, SGX exits the enclave and generates a $\tt{SIGILL}$ signal.
Gracefully recovering from the failure requires re-entering the
enclave at a different program point. Due to R5, this is disallowed by
SGX. In \codename, either \dr or the application can invoke illegal 
instructions, which may create unanticipated exits from the enclave.

\codename has three ways to handle them: (a) entirely delegate the
instruction outside the enclave (e.g., file, networking, and timer
operations); (b) execute the instruction outside the enclave while
explicitly updating the in-enclave state (e.g., thread operations,
signal handling); or (c) completely simulate instruction inside the
enclave. 
 
\codename changes \dr logic to convert such illegal instruction to
stubs that either delegate or emulate the functionality. For the
target application, whenever \codename observes an illegal
instructions in the code cache, it replaces the instruction with a
call to the \codename syscall handler function. 

Syscalls are a special case---they access process memory for
input-output parameters and error codes. Since enclaves do not allow
this, for delegating the syscall outside the enclave, \codename
creates a copy of input parameters from private memory to public
memory. This includes simple value copies as well as deep copies of
structures. The OS then executes the syscall and generates results in
public memory. Post-call, \codename copies back the explicit results
and error codes to private memory. 

Memory copies alone are not sufficient. For example, when loading a
library, the application uses $\tt{dl\_open}$ which in turn calls
$\tt{mmap}$. When we execute the $\tt{mmap}$ call outside the enclave,
the library is mapped in the untrusted public address space of the
application. However, we want it to be mapped privately inside the
enclave. As another example, consider when the enclave wants to create
a new thread local storage (TLS) segment. If \codename executes the
system call outside the enclave, the new thread is created for the \dr
runtime instead of the target application. Thus, when a syscall
implicitly changes application state, \codename has to explicitly
propagate those changes inside the enclave.

Alternatively, \codename selectively emulates some syscalls inside
the enclave. For example, $\tt{arch\_prctl}$ is used to read the FS
base, \codename substitutes it with a $\tt{rdfsbase}$ instruction and
executes it inside the enclave. We outline the details of other
syscall subsystems that are fully or partially emulated by \codename
in Sections~\ref{sec:memmgmt}, \ref{sec:threading}, \ref{sec:synch},
\ref{sec:sighand}.

\codename resumes execution only after the syscall state has been
completely copied inside the enclave. This allows it to employ
sanitization of OS results before using it. Further, all the
subsequent execution is strictly over private memory to avoid TOCTOU
attacks. \subsection{Memory Management}
\label{sec:memmgmt}

For syscalls that change process virtual address layout, \codename has
to explicitly reflect their changes inside the enclave. First, this is
not straightforward. Due to R1-R4, several layout configurations are
not allowed for enclave virtual memory (e.g., changing memory
permissions). Second, \codename does not trust the information
provided by the OS (e.g., via $\tt{procmap}$).
 
To address these challenges, \codename maintains its own
$\tt{procmap}$-like structure. Specifically, \codename keeps its own
view of the process virtual memory inside the enclave, tracks the
memory-related events, and updates the enclave state. For example,
after $\tt{mmap}$ call succeeds outside the enclave, \codename
allocates and records the new virtual address located inside the
enclave.

Further, \codename synchronizes the two-copies of memories to maintain
execution semantics. For example, after $\tt{mmap}$, \codename creates
a new memory mapping inside the enclave and then copies the content of
the $\tt{mmap}$-ed memory inside the enclave. On subsequent changes to
$\tt{mmap}$ed-memory, \codename updates the non-enclave memory via a write.
This is done whenever the application either unmaps the memory or
invokes sync or $\tt{fsync}$ system call.

With mediation over memory management, \codename transparently
side-steps SGX restrictions. When application makes requests that are
not allowed in SGX (e.g., changing memory permissions), \codename
replaces it with a sequence of valid SGX operations that achieve the
same effect (e.g., move the content to memory which has the required
permissions). Subsequently, when the application binary accesses
memory, \codename can pre-emptively replace the addresses to access
the correct in-enclave copy of the memory. This allows us to safely
and transparently mimic disallowed behavior inside the enclave.

\codename does not blindly replicate OS-dictated memory layout changes
inside the enclave. It first checks if the resultant layout will
violate any security semantics (e.g., mapping a buffer to
zero-address). It proceeds to update enclave layout and memory content
only if these checks succeed. To do this, \tool keeps its metadata in
private  memory. \begin{figure}[tbp]
\centering
\includegraphics[width=0.3\textwidth]{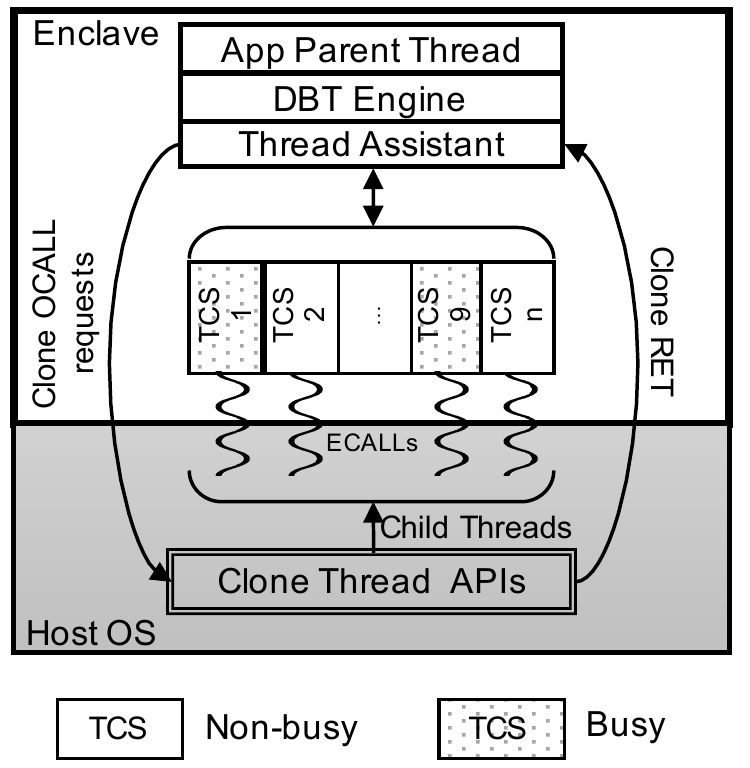}
\caption{
Design for multi-threading in \codename. 
}
\label{fig:thread-design}
\end{figure}

\subsection{Multi-threading} 
\label{sec:threading}

SGX requires the application to pre-declare the maximum number of
threads before execution (R2). Further, it does not allow the enclave
to resume at arbitrary program points or execution contexts (R5). This
creates several compatibility and security challenges in \codename.

\dr and the target application share the same thread, but they have
separate TLS segment for cleaner context switch. \dr keeps the threads
default TLS segment for the target application and creates a new TLS
segment for itself at a different address. \dr switches between these
two TLS segments by changing the segment register---\dr uses
$\tt{gsbase}$, application uses $\tt{fsbase}$.
SGX allocates one TLS segment per enclave thread. SGX uses the same
mechanism as \dr (i.e., changing the segment) to maintain a shadow TLS
segment for itself when executing enclave code.

\paragraph{Multiplexing Base Registers.}
When we attempt to run \dr inside SGX, there are not enough registers
to save three offsets (one for \dr, one for SGX, one for the
application). To circumvent this limitation, \codename adds two TLS
segment fields to store $\tt{fsbase}$ and $\tt{gsbase}$ register
values. We use these TLS segment fields to save and restore pointers
to the segment base addresses. This allows us to maintain and switch
between three clean TLS segment views per thread.

\paragraph{Primary-secondary TLS Segment Design.}
Since \codename is in-charge of maintaining the view of
multiple-threads, it has to switch the TLS segment to a corresponding
thread every time the execution enters or exits the enclave. We
simplify these operations with a primary-secondary TLS segment design.
\codename adds a new field to the SGX thread data structure---a flag
to indicate if the TLS segment is the primary or not.

\codename marks the default first TLS segment created by SGX as the
primary. To do this, it sets the flag of the TLS segment when the
execution enters the enclave for the first time after creation. All
the subsequent TLS segment, if created, are marked as secondary. If
the flag is false, the base value stores the pointer to the addresses
of the primary TLS segment. Otherwise, it points to the secondary TLS
segment required to execute \dr. With this mechanism, upon enclave
entry, \codename circulates through the TLS segment pointers until it
finds the addresses for the primary TLS segment.

\paragraph{Dynamic Threading.}
Since the number of TCS entries is fixed at enclave creation time, 
the maximum number of threads supported is capped.
\codename multiplexes the limited TCS entries, as shown in 
Figure~\ref{fig:thread-design}. When an application wants to create a
new thread  (e.g., via clone),  \codename first checks if there is a
free TCS slot. If it is the case, it performs an \ocall to do so
outside the enclave. Otherwise, it busy-waits until a TCS slot is
released. Once a TCS slot is available, the \ocall creates a new
thread outside the enclave.  After finishing thread creation, the
parent thread returns back to the enclave and resumes execution. The
child thread explicitly performs an \ecall to enter the enclave and
\dr resumes execution for the application's child thread.

For all threading operations, \codename ensures transparent
context switches to preserve binary compatibility as intended by \dr.
For security, \codename creates and stores all thread-specific context
either inside the enclave or SGX's secure hardware-backed storage at
all times. It does not use any OS data structures or addresses for
thread management. \subsection{Thread Synchronization}
\label{sec:synch}

SGX provides basic synchronization primitives (e.g., SGX mutex and
conditional locks) backed by hardware locks. But they can only be used
for enclave code. Thus, they are semantically incompatible with the
lock mechanisms used by \dr or legacy applications which use OS locks.
For example, \dr implements a sophisticated mutex lock using
$\tt{futex}$ syscall, where the lock is kept in the kernel memory.
Supporting this requires trusting the OS for sharing locks. Given R1
and R3, SGX does not offer any memory sharing model, making it
impossible to support $\tt{futex}$es.

\begin{figure}[tbp]
\centering
\includegraphics[width=0.5\textwidth]{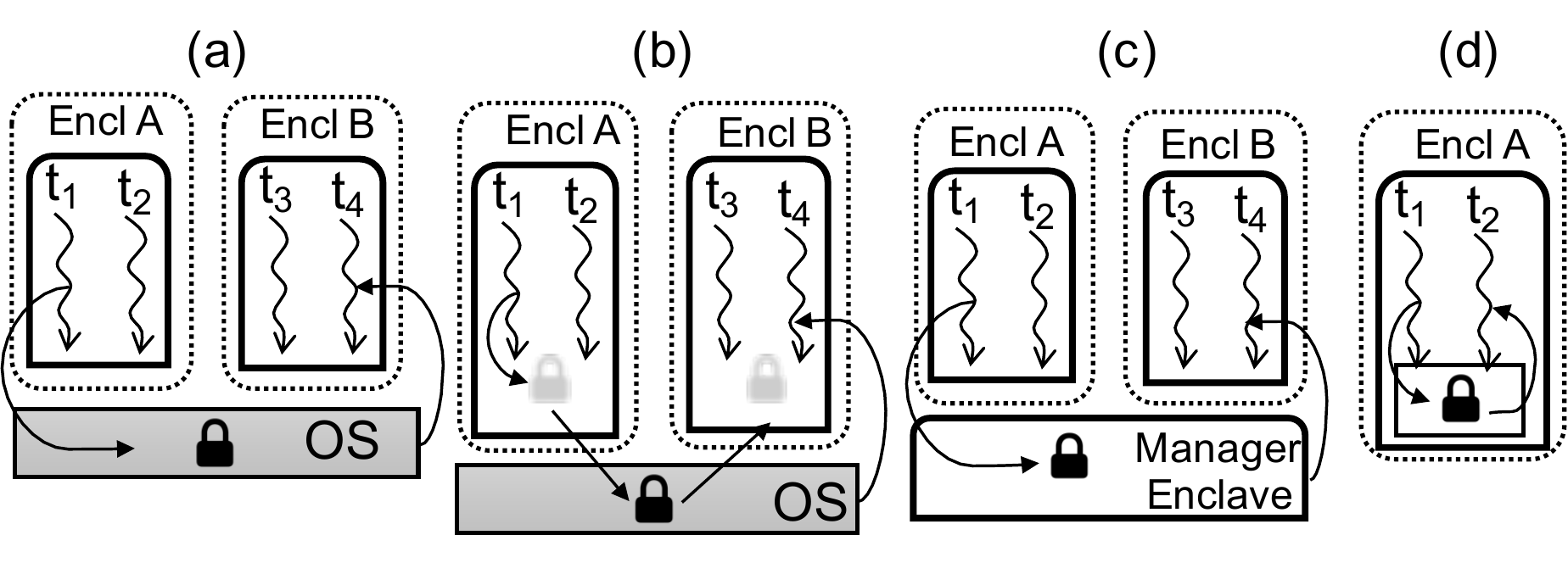}
\caption{
Design choices for lock synchronization. 
(a) Traditional Futex. 
(b) Two-copy design with futex in public memory. 
(c) Dedicated lock manager in a separate enclave. 
(d) \codename case where the threads are in the same enclave. 
}
\label{fig:sync-design}
\end{figure}

\paragraph{Need for a Lock Manager.}
A naive design would be to maintain a shadow $\tt{futex\ variable}$ in public
memory, such that it is accessible to the enclave(s) and the OS.
However, the OS can arbitrarily change the lock state and attack the
application.
As an alternative, we can employ a two-copy mechanism for locks. The
enclave can keep the lock in private memory. When it wants to
communicate state change to the OS, \codename can tunnel a futex OCALL
to the host OS. There are several problems even with this approach. 
Threads inside the enclave may frequently update the locks in private
memory. Polling or accessing the futex outside the enclave (including
the kernel and the untrusted part of the enclave) requires the latest
state of lock every time there is an update in private memory. 
This creates an opportunity for the OS to launch TOCTOU attack. 
Even without TOCTOU attacks, it is challenging to synchronize the two
copies in benign executions. Specifically, private lock states can be
changed while the public state is being updated. This results in
threads inside and outside the enclave with inconsistent views of the
same lock (i.e., private and public copy). The more frequent the
local updates, the higher is the probability of such 
inconsistencies.
In general, the only way to avoid such race condition bugs is to use
locks for synchronizing the private and public state of the
enclave locks.
This is impossible with SGX because it does not support secure memory
sharing between the OS and the enclave(s).
Figure~\ref{fig:sync-design} shows the schematics of design choices
for implementing synchronization primitives.

\paragraph{\codename Lock Manager.}
Given the futex-usage of \dr, we identify that we can avoid sharing an
enclave's lock directly with the OS or other enclaves. The \dr usage
of futexes can be replaced with a simpler primitive such as spinlocks
to achieve the same functionality. Specifically, we implement a lock
manager in \codename. We use the hardware spinlock exposed by SGX to
do this securely and efficiently inside the enclave. In \codename we
invoke our lock manager implementation wherever \dr tries to use
futexes. The other instance of futex usage is in the application
binaries being executed with \codename. To handle those cases, when
\codename loads application binary into the code cache, it replaces
thread-related calls (e.g., $\tt{pthread\_cond\_wait}$) with stubs to
invoke our lock manager to use safe synchronization. \subsection{Signal Handling}
\label{sec:sighand}

\begin{figure*}[tbp]
\centering
\includegraphics[width=0.7\textwidth]{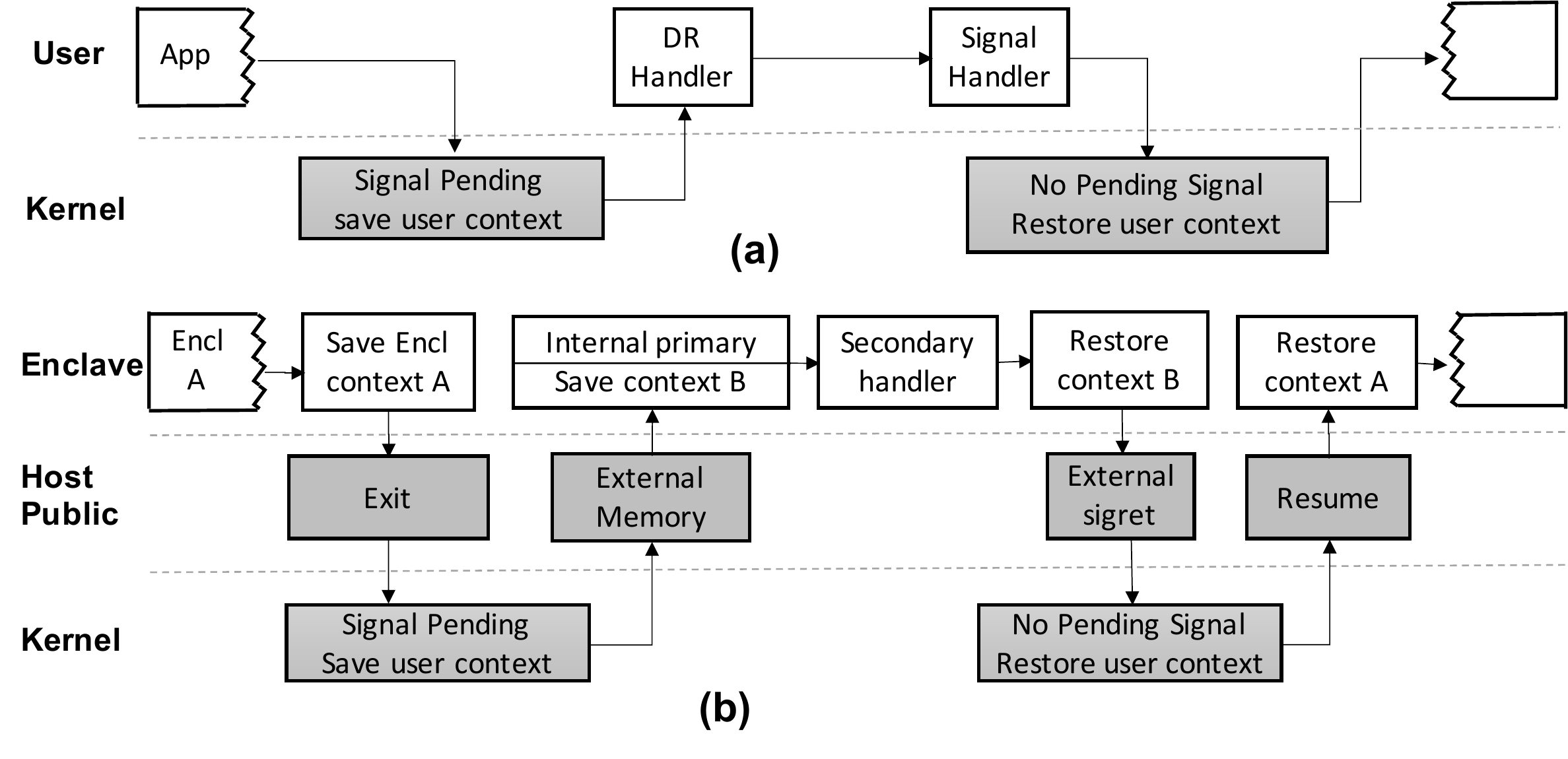}
\caption{
(a) Original signal design in \dr. 
(b) Design for signal design in \codename. 
}
\label{fig:signal-design}
\end{figure*}
 
\codename cannot piggyback on the existing signal handling mechanism
exposed by the SGX, due to R5. Specifically, when \dr executes inside
the enclave, the \dr signal handler needs to get description of the
event to handle it (Figure~\ref{fig:signal-design}(a)). However,
Intel's SGX platform software removes all such information when it
delivers the signal to to the enclave.
This breaks the functionality of programmer-defined handlers to
recover from well known exceptions (divide by zero). Further, any
illegal instructions inside the enclave generate exceptions, which are
raised in form of signals. Existing binaries may not have handlers for
recovering from instructions illegal in SGX.

SGX allows entering the enclave at fixed program points and context. 
Leveraging this, \codename employs a {\em primary} signal handler that
it registers with SGX. For any signals generated for \dr or the
application, we always enter the enclave via the primary handler and
we copy the signal code into the enclave. We then use the primary as a
trampoline to route the control to the appropriate {\em secondary}
signal handler inside the enclave, based on the signal code. 
At a high-level, we realize a virtualized trap-and-emulate signal handling design. We use SGX signal
handling semantics for our primary. For the secondary, we setup and
tear down a separate stack to mimic the semantics in the software. The
intricate details of handling the stack state at the time of such context 
switched are elided here. 
Figure~\ref{fig:signal-design}(b) shows a schematic of our design. 

\paragraph{Registration.}
Original \dr code and the application binary use $\tt{sigaction}$ to
register signal handlers for itself. We refer to them as secondary
handlers. In \codename, first we change \dr logic to register only the
primary signal handler with SGX. We then record the \dr and
application registrations as secondary handlers.
This way, whenever SGX or the OS wants to deliver the signal for the
enclave, SGX directs the control to our primary handler inside the 
enclave\footnote{Vanilla SGX PSW does not provide an API which allows
the enclave to register for signals, we have changed the PSW to
support this.}. Since this is a pre-registered handler, SGX allows it.
The primary handler checks the signal code and explicitly routes
execution to the secondary. 

\paragraph{Delivery.}
A signal may arrive when the execution control is inside the enclave
(e.g., timer). In this case, \codename executes a primary signal
handler code that delivers the signal to the enclave. However, if the
signal arrives when the CPU is in a non-enclave context, SGX does not
automatically invoke the enclave to redirect execution flow. To force
this, \codename has to explicitly enter the enclave. But it can only
enter at a pre-registered program point with a valid context (R5).
Thus we first wake up the enclave at a valid point (via \ecall) and
copy the signal code. We then simulate the signal delivery by setting
up the enclave stack to execute the primary handler.

\paragraph{Exit.}
After executing their logic, handlers use $\tt{sigreturn}$ instruction
for returning control to the point before the signal interrupted the
execution. When \codename observes this instruction in the secondary
handler it has to simulate a return back to the primary handler
instead. The primary handler then performs its own real 
$\tt{sigreturn}$. SGX then resumes execution from the point before the
signal was generated.
  \section{Implementation}
\label{sec:impl}

We implement \codename by using \dr. We run \dr inside an enclave 
with the help of standard Intel SGX development kit that includes 
user-land software (SDK v2.1), platform software (PSW v2.1), and a
Linux kernel driver (v1.5). We make a total of $9667$ \loc software
changes to \dr and SDK infrastructure. We run \codename on an
unmodified hardware that supports SGX v1.

\codename design makes several changes to \dr core (e.g., memory
management, lock manager, signal forwarding). When realizing our
design, we address three high-level implementation challenges. Their
root cause is the way Intel SDK and PSW expose hardware features and
what \dr expects. There are several low-level challenges that we do
not discuss here for brevity.

\paragraph{Self-identifying Load Address.}
\dr needs to know its location in memory mainly to avoid using its own
address space for the application. In \codename, we use a
$\tt{call-pop}$ mechanism to self-identify our location in memory.
These contiguous instructions, allow us to dynamically retrieve our own 
address. Specifically, we align this address and decrease it at a
granularity of page-size to compute our start-address in memory.

\paragraph{Insufficient Hardware Slots.}
By default, SGX SDK and PSW assume two SSA frames, which are
sufficient to handle most of the nested signals. Since \codename
design needs one SSA frame for itself, we increase the SSA frames to
three to ensure we can handle the same set of nested signals as SGX. 
The SGX specification allows this by changing the $\tt{NSSA}$ field in
our PSW implementation.

\paragraph{Preserving Execution Contexts.}
For starting execution of a newly created thread, \codename invokes
a pre-declared \ecall to enter the enclave. This is a nested ECALL,
which is not supported by SGX SDK. 
To allow it, we modify the SDK to facilitate the entrance of child
threads and initialize the thread data structure for it. Specifically,
we check if the copy of thread arguments inside the enclave matches the
ones outside before resuming thread execution. We save specific
registers so that the thread can exit the enclave later. 
Note that the child thread has its own execution path differentiating
from the parent one, \codename hence bridges its return address to the
point in the code cache that a new thread always starts.
After the thread is initialized, we explicitly update \dr data
structures to record the new thread (e.g., the TLS base for
application libraries) This way, \dr is aware of the new thread and
can control its execution in the code cache. 

\paragraph{Propagating Implicit Changes \& Metadata.}
Thread uses $\tt{exit/exit\_group}$ syscall for terminating itself.
Then the OS zeros out the child thread ID ($\tt{ctid}$). In
\codename, we explicitly create a new thread inside the enclave, so
we have to terminate it explicitly by zeroing out the pointers to the
IDs. Further, we clean up and free the memory associated with each
thread inside and outside the enclave.
 \section{Evaluation}
\label{sec:eval}

We primarily evaluate the compatibility of \tool and highlight the
advantages gained due to binary compatibility. We further provide  a
TCB breakdown of \tool and point out the performance ramifications,
which are common and comparable to state-of-the-art other approaches.

\paragraph{Setup.}
All our experiments are performed on a Lenovo machine with SGX v$1$
support, $128$ MB EPC configured in the BIOS of which $96$ MB is
available for user-enclaves, $12$ GB RAM, $64$ KB L1, $256$ KB L2,
$4096$ KB L3 cache, $3.4$GHz processor speed. Our software setup
comprises Ubuntu version $16.04$, Intel SGX SDK v$2.1$, PSW v$2.1$,
driver v$1.5$, gcc v$5.4.0$, \dr v$6.2.17$. All performance statistics
reported are  the geometric mean over $5$ runs.

\subsection{Compatibility}
\label{sec:compat}

\paragraph{Selection Criteria.}
We select $310$ binaries that cover an extensive set of benchmarks,
utilities, and large-scale applications. They comprise  commonly used
evaluations target for \dr and enclave-based
systems~\cite{graphene-sgx, scone, privado, besfs, podarch, drmem,
dbt-tsx} that we surveyed for our study. Further, they represent a mix
of memory, CPU, multi-threading, network, and file I/O workloads.
Our $69$ binaries are from micro-benchmarks targets: $29$ from
SPEC2006 (CPU), $1$ from IOZone (I/O) v$3.487$, $9$ from FSCQ v$1.0$
(file API), $21$ from HBenchOS v$1.0$ (system stress-test), and $9$
from Parsec-SPLASH2 (multi-threading). We run $12$ binaries from $3$
real-world applications---cURL v$7.65.0$ (server-side utilities),
SQLite v$3.28.0$ (database), Memcahed v$1.5.20$ (key-value store), and
$9$ applications from Privado (secure ML framework). We test $229$
Linux utilities from our system's $\tt{/bin}$ and $\tt{/usr/bin}$
directories.

\paragraph{Porting Efforts.}
For benchmarks and applications, we download the source code and
compile them with default flags required to run them natively on our
machine. We directly use the existing binaries for Linux utilities. We
test the same binaries on native hardware, with \dr, and with
\codename. Thus, we do not change the original source-code or the
binaries. We test the target binaries on native Linux and on vanilla
\dr. $278/310$ of targets execute successfully with these baselines. 

The remaining $32$ binaries either use unsupported devices (e.g.,
NTFS) or do not run on our machine. So we discard them from our
\codename experiments, since vanilla \dr also does not work on them. 
Of the remaining $278$ binaries that work on the baselines, \codename
has support for the system calls used by $203$ of these. We support
{\em all} of these and directly execute them with \codename, with zero
porting effort.

\begin{table}[tbp]
    \centering
    \resizebox{0.5\textwidth}{!}{
    \begin{tabular}{@{}lrrrrrc@{}}
    \toprule
     \multicolumn{1}{c}{\multirow{2}{*}{\textbf{\begin{tabular}[c]{@{}c@{}}Subsystem\end{tabular}}}} & \multicolumn{1}{c}{\multirow{2}{*}{\textbf{Total}}} & \multicolumn{1}{c}{\multirow{2}{*}{\textbf{Impl}}} & \multicolumn{3}{c}{\textbf{Implementation}} & \multicolumn{1}{c}{\textbf{Covered}}\\
     \cmidrule(l){4-6} \cmidrule(l){7-7}
    \multicolumn{1}{c}{} & \multicolumn{1}{c}{} & \multicolumn{1}{c}{} & \multicolumn{1}{c}{\textbf{Del}} & \multicolumn{1}{c}{\textbf{Emu}} & \multicolumn{1}{c}{\textbf{P.Emu}} & \multicolumn{1}{c}{\textbf{DR + Binaries}}\\ 
    \midrule
     Process & 12 & 8 & 4 & 2 & 2 & 3 \\
     Filename based & 37 & 25 & 25 & 0 & 0 & 16 \\
     Signals & 12 & 7 & 3 & 4 & 0 & 6 \\
     Memory & 18 & 10 & 6 & 0 & 4 & 4 \\
     Inter process communication & 12 & 4 & 4 & 0 & 0 & 0 \\
     File descriptor based & 65 & 53 & 48 & 0 & 5 & 30 \\
     File name or descriptor based & 19 & 9 & 9 & 0 & 0 & 5 \\
     Networks & 19 & 17 & 15 & 0 & 2 & 15 \\
     Misc & 124 & 79 & 79 & 0 & 0 & 36 \\
     \midrule
     
    {Total} & 318 & 212 & 193 & 6 & 13 & 115 \\ \bottomrule
    \end{tabular}
    }
    \caption{\codename syscall support.
    Column $2-3$: total Linux system calls and support in \tool.
    Column $4-6$: syscalls implemented by full delegation, full emulation, and partial emulation. 
    Column $7$: syscalls tested in \tool.}
    \label{tab:syscall}
    \end{table} \begin{figure*}[tbp]
    \centering
    \begin{minipage}[h]{0.32\textwidth}
    \includegraphics[width=\textwidth]{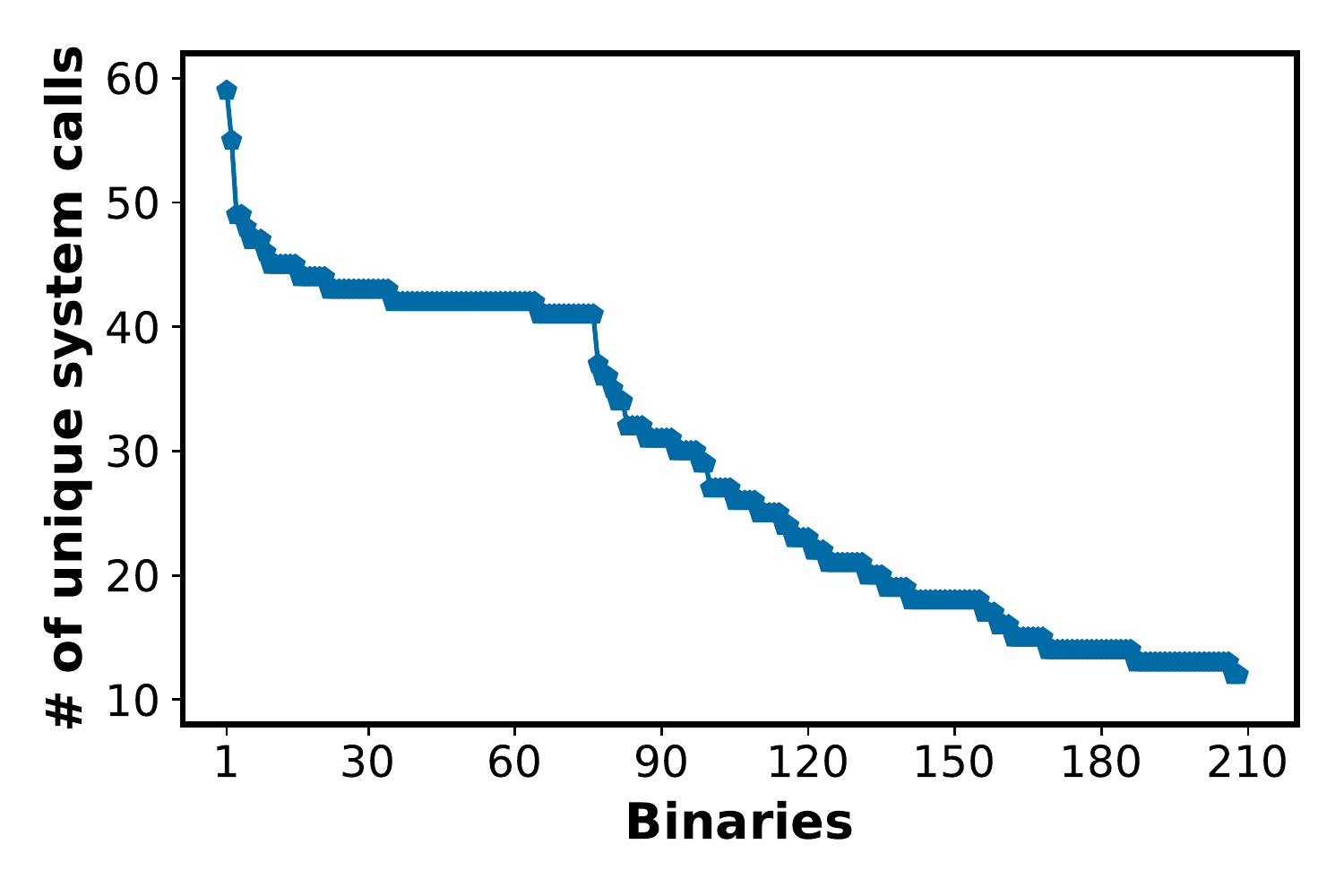}
    \subcaption{Number of unique syscalls}
    \label{fig:syscount}
    \end{minipage}
    \begin{minipage}[h]{0.32\textwidth}
    \includegraphics[width=\textwidth]{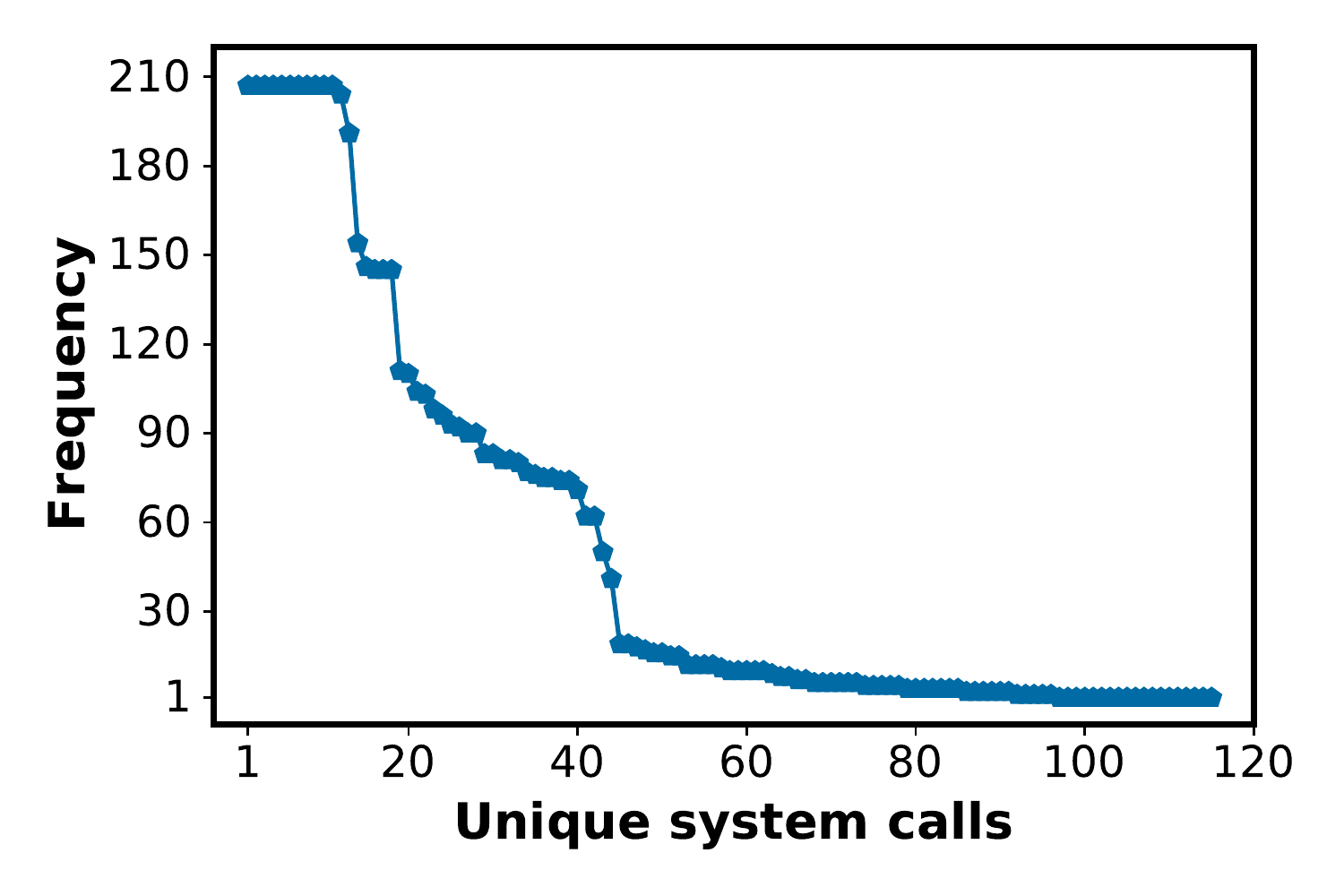}
    \subcaption{Syscall frequency}
    \label{fig:sysfrequency}
    \end{minipage}
    \caption{
    System calls statistics 
over all $208$ binaries. 
    (a) Unique syscalls for each binary; and (b) frequency per syscall. }
    \label{fig:apps-syscall-frequency}
\end{figure*} \begin{table}[tbp]
\centering
\resizebox{0.5\textwidth}{!}{
\begin{tabular}{@{}lrrrrrrrrrrr@{}}
\toprule
\multicolumn{1}{c}{\multirow{2}{*}{\textbf{Function}}} & \multicolumn{3}{c}{\textbf{Trusted}} & \multicolumn{5}{c}{\textbf{Untrusted}} \\ \cmidrule(l){2-4} \cmidrule(l){5-9} 
\multicolumn{1}{c}{} & \textbf{SDK+PSW} & \textbf{DR} & \textbf{Total} & \textbf{SDK+PSW} & \textbf{DR} & \textbf{Driver} & \textbf{Host}& \textbf{Total} & \multicolumn{1}{c}{} \\ \midrule
Original & 147928 & 129875 & 277803 & 49838 & 66629 & 2880 & 1769 & 121116 \\
Loader & 69 & 1604 & 1673 & 27 & 89 & N/A & 332 & 448 \\ 
MM & 46 & 2241 & 2287 & 44 & 0 & N/A & 0 & 44 \\ 
Syscalls & 0 & 1801 & 1801 & 0 & 0 & N/A & 1432 & 1432 \\ 
Instr & 0 & 45 & 45 & 0 & 0 & N/A & 26 & 26 \\ 
TLS & 18 & 60 & 78 & 18 & 0 & N/A & 0 & 18 \\ 
Signals & 201 & 236 & 437 & 136 & 0 & N/A & 0 & 136 \\ 
Threading & 389 & 393 & 782 & 130 & 0 & N/A & 157 & 287 \\ 
Sync & 0 & 173 & 173 & 0 & 0 & N/A & 0 & 0 \\  \bottomrule
\end{tabular}
}
\caption{Breakdown of \codename TCB.}
\label{tab:tcb}
\end{table} 

\paragraph{System Call Support \& Coverage.}
\codename supports a total of $212/318$ ($66.66\%$) syscalls exposed
by the Linux Kernel. We emulate $6$ syscalls purely inside the enclave
and delegate $193$ of them via OCALLs. For the remaining $13$, we use
partial emulation and partial delegation. Table~\ref{tab:syscall}
gives a detailed breakdown of our syscall support.
Syscall usage is not uniform across frequently used applications and
libraries~\cite{syscall-api-eurosys}. Hence we empirically evaluate
the degree of expressiveness supported by \codename. For all of the
$278$ binaries in our evaluation, we observe total $121$ unique
syscalls, \codename supports $115$ of them.
Table~\ref{tab:syscall} shows the syscalls supported by \codename and
their usage in our benchmarks and real-world applications (See
Appendix~\ref{sec:appx-utilities}).
Figure~\ref{fig:syscount} and ~\ref{fig:sysfrequency} show the
distribution of unique syscalls and their frequency as observed over
binaries supported by \codename. Thus, our empirical study shows that
\codename supports $115/121$ ($95.0\%$) syscall observed in our
dataset of binaries.
To support $212$ syscalls, we added $3233$ \loc (10 \loc per syscall
on average). In the future, \codename can be extended to
increase the number of supported syscalls. Interested readers can
refer to Appendix~\ref{sec:appx-utilities} for more details.
\tool handles $31$ out of the $32$ standard signals defined
for Linux. It does not handle \texttt{SIGPROF} presently because
the original \dr has no support for it natively.

There are $72$ binaries that use syscalls that are presently
unsupported in \tool. $3$ from HBenchOS fail because \codename does
not support $\tt{fork}$.
$1$ from SPEC2006 fails because SGX has insufficient virtual memory.
$68$ are Linux utilities of which $39$ are multi-process, $10$ use
unsupported ioctls, $12$ use unsupported signals, $1$ fails with the
same reason as the one from SPEC2006, and $6$ use other unsupported
system calls.

\begin{table*}[tbp]
\centering
\resizebox{0.7\textwidth}{!}{
\begin{tabular}{@{}llrrrrrrSSS@{}}
\toprule
\multicolumn{1}{c}{\multirow{2}{*}{\textbf{\begin{tabular}[c]{@{}c@{}}Suite\\ Name\end{tabular}}}} &
  \multicolumn{1}{c}{\multirow{2}{*}{\textbf{\begin{tabular}[c]{@{}c@{}}Benchmark\\ /Application\\ Name\end{tabular}}}} &
  \multicolumn{2}{c}{\textbf{Compile Stats}} &
  \multicolumn{4}{c}{\textbf{Runtime Stats}} &
  \multicolumn{2}{c}{\textbf{Time (sec)}} &
  \multicolumn{1}{c}{\multirow{2}{*}{\textbf{\begin{tabular}[c]{@{}c@{}}Overhead\\ (in \%)\end{tabular}}}} \\ \cmidrule(l){3-4} \cmidrule(l){5-8} \cmidrule(l){9-10}  
\multicolumn{1}{c}{} &
  \multicolumn{1}{c}{} &
  \multicolumn{1}{c}{\textbf{LOC}} &
  \multicolumn{1}{c}{\textbf{Size}} &
  \multicolumn{1}{c}{\textbf{\begin{tabular}[c]{@{}c@{}}Out\\ Calls\end{tabular}}} &
  \multicolumn{1}{c}{\textbf{\begin{tabular}[c]{@{}c@{}}Sys\\ Calls\end{tabular}}} &
  \multicolumn{1}{c}{\textbf{\begin{tabular}[c]{@{}c@{}}Page\\ Faults\end{tabular}}} &
  \multicolumn{1}{c}{\textbf{\begin{tabular}[c]{@{}c@{}}Ctx\\ Swt\end{tabular}}} &
  \multicolumn{1}{c}{\textbf{DR}} &
  \multicolumn{1}{c}{\textbf{Ratel}} &
  \multicolumn{1}{c}{} \\ \midrule
\multirow{11}{*}{\ver{SPEC CINT2006}} 
 & astar & 4280 & 56 KB & 26561 & 618555 & 271544 & 181 & 8.77 & 10.71 & 18.13 \\
 & bzip2 & 5734 & 73 KB & 26048 & 618115 & 443869 & 238 & 21.74 & 34.49 & 36.96 \\
 & gobmk & 157650 & 4.4 MB & 26594 & 618629 & 272957 & 150 & 1.73 & 4.37 & 60.43 \\
 & hmmer & 20680 & 331 KB & 26144 & 629203 & 271106 & 139 & 0.98 & 0.85 & -15.45 \\
 & sjeng & 10549 & 162 KB & 26606 & 618638 & 2049404 & 382 & 5.08 & 4.57 & -11.20 \\
 & libquantum & 2611 & 51 KB & 25969 & 618004 & 271071 & 150 & 0.50 & 0.47 & -7.81 \\
 & h264ref & 36097 & 602 KB & 27033 & 619033 & 272100 & 284 & 18.25 & 34.83 & 47.60 \\
 & omnetpp & 26652 & 871 KB & 26961 & 618990 & 271813 & 151 & 2.47 & 2.72 & 9.20 \\
 & Xalan & 267376 & 6.3 MB & 28121 & 620185 & 273953 & 198 & 5.53 & 4.05 & -36.64 \\
 & gcc & 385783 & 3.8 MB & 25758 & 656241 & 56201 & 454 & 12.85 & 6.30 & -103.97 \\
 & gromac & 87921 & 1.1 MB & 26783 & 654600 & 55019 & 633 & 2.85 & 4.79 & 40.50 \\
 \midrule
\multirow{13}{*}{\ver{SPEC CFP2006}}
 & leslie3d & 2983 & 177 K & 26831 & 618865 & 271723 & 204 & 18.80 & 21.63 & 13.05 \\
 & milc & 9580 & 150 KB & 32551 & 624587 & 271506 & 192 & 13.05 & 22.41 & 41.76 \\
 & namd & 3892 & 330 KB & 28550 & 620582 & 271665 & 173 & 18.86 & 19.41 & 2.85 \\
 & cactusADM & 60235 & 819 KB & 27619 & 619634 & 370217 & 190 & 5.34 & 7.78 & 31.35 \\
 & calculix & 105123 & 1.8 MB & 27319 & 629243 & 313313 & 174 & 3.06 & 3.90 & 21.44 \\
 & dealII & 94458 & 4.3 MB & 26858 & 618872 & 273471 & 240 & 24.68 & 24.73 & 0.20 \\
 & GemsFDTD & 4883 & 440 KB & 25207 & 617226 & 366889 & 177 & 5.44 & 2.08 & -161.25 \\
 & povray & 78684 & 1.2 MB & 29082 & 621108 & 272267 & 166 & 4.42 & 4.72 & 6.38 \\
 & soplex & 28282 & 507 KB & 26880 & 618909 & 271861 & 164 & 2.07 & 2.08 & 0.31 \\
 & specrand & 54 & 8.7 KB & 25863 & 617897 & 270924 & 150 & 0.35 & 0.27 & -27.85 \\
 & specrand & 54 & 8.7 KB & 25863 & 617897 & 270990 & 164 & 0.34 & 0.33 & -2.35 \\
 & tonto & 107228 & 4.6 MB & 30562 & 622574 & 273669 & 186 & 6.89 & 6.65 & -3.53 \\
 & zeusmp & 19030 & 280 KB & 27163 & 619201 & 1755130 & 434 & 20.65 & 52.03 & 60.32 \\
 \midrule
 \multirow{5}{*}{\ver{IOZONE}}
 & read\/re-read & 26545 & 1.1 MB & 27254 & 622791 & 23785 & 1215 & 0.88 & 0.89 & 1.35 \\
 & random read\/write & 26545 & 1.1 MB & 27376 & 622913 & 23744 & 844 & 0.88 & 1.09 & 19.41 \\
 & read backward & 26545 & 1.1 MB & 27431 & 622968 & 23854 & 1159 & 0.84 & 1.38 & 39.22 \\
 & fwrite\/re-fwrite & 26545 & 1.1 MB & 27212 & 622750 & 24317 & 581 & 0.87 & 0.86 & -0.93 \\
 & iozone fread\/re-fread & 26545 & 1.1 MB & 27223 & 622760 & 23742 & 374 & 0.86 & 0.65 & -31.90 \\
  \midrule
\multirow{9}{*}{\ver{FSCQ}}
 & fscq large file & 383 & 25 KB & 25889 & 1165892 & 270914 & 168 & 0.47 & 3.41 & 86.22 \\
 & fscq small file & 161 & 19 KB & 26352 & 929795 & 270959 & 181 & 0.34 & 0.17 & -104.82 \\
 & fscq write file & 74 & 18 KB & 262015 & 930226 & 270867 & 143 & 0.31 & 0.13 & -142.19 \\
 & multicreatewrite & 20 & 11 KB & 65721 & 969595 & 270969 & 248 & 0.38 & 0.83 & 54.59 \\
 & multiopen & 14 & 9.8 KB & 225719 & 1129593 & 270842 & 452 & 0.57 & 2.44 & 76.64 \\
 & multicreate & 18 & 9.9 KB & 55720 & 949625 & 270866 & 212 & 0.31 & 0.67 & 53.59 \\
 & multiwrite & 16 & 9.9 KB & 35720 & 939594 & 270864 & 152 & 0.23 & 0.30 & 22.52 \\
 & multicreatemany & 19 & 11 KB & 45729 & 959605 & 271034 & 198 & 0.36 & 0.77 & 53.63 \\
 & multiread & 17 & 9.9 KB & 325721 & 1229595 & 270901 & 589 & 0.62 & 3.70 & 83.20 \\ 
  \midrule
\multirow{9}{*}{\ver{PARSEC-SPLASH2}}
& water\_nsquare & 2885 & 46 KB & 27109 & 622992 & 25093 & 199 & 0.94 & 0.88 & -6.44 \\
& water\_spatial & 3652 & 46 KB & 27171 & 622991 & 61992 & 164 & 1.05 & 2.31 & 54.69 \\
& barnes & 4942 & 46 KB & 26801 & 622748 & 28191 & 287 & 0.85 & 1.02 & 16.59 \\
& fmm & 7611 & 64 KB & 27218 & 622909 & 62025 & 58 & 0.97 & 2.18 & 55.72 \\
& raytrace & 200091 & 92 KB & 27291 & 623175 & 65323 & 194 & 1.23 & 2.59 & 52.32 \\
& radiosity & 21586 & 230 KB & 27609 & 623118 & 26162 & 139 & 4.30 & 5.82 & 26.01 \\
& ocean\_cp & 10519 & 81 KB & 27234 & 622990 & 25480 & 86 & 1.19 & 1.05 & -12.56 \\
& ocean\_ncp & 6275 & 65 KB & 27052 & 622948 & 25362 & 91 & 1.03 & 1.08 & 4.71 \\
& volrend & 27152 & 271 KB & 27309 & 623167 & 25082 & 178 & 0.75 & 0.88 & 14.56 \\
\midrule
\multirow{12}{*}{\ver{Applications}}
 & SQLite (10 K keys) & 140420 & 1.3 MB & 400548 & 1818195 & 272323 & 1261 & 5.82 & 6.94 & 16.19 \\
 & cURL (10 MB) & 22064 & 30 KB & 35897 & 940802 & 272552 & 1031 & 1.78 & 1.17 & -51.54 \\
 & Memcached (100K) & 44921 & 795 KB & 1021241 & 1118691 & 540649 & 104765 & 5.99 & 9.46 & 36.68 \\
 & densenetapp & 12551 & 32 MB & 27826 & 616749 & 123894 & 354 & 7.25 & 12.90 & 43.77 \\
 & lenetapp & 230 & 313 KB & 26029 & 616411 & 21166 & 362 & 0.49 & 0.26 & -92.43 \\
 & resnet110app & 9528 & 110 MB & 27238 & 696291 & 23716 & 200 & 1.98 & 2.45 & 18.82 \\
 & resnet50app & 2826 & 98 MB & 26591 & 616291 & 136139 & 274 & 7.64 & 11.81 & 35.36 \\
 & resnext29app & 1753 & 132 MB & 26410 & 616728 & 187284 & 411 & 11.38 & 16.33 & 30.33 \\
 & squeezenetapp & 914 & 4.8 MB & 26258 & 616290 & 23001 & 252 & 1.22 & 1.11 & -9.83 \\
 & vgg19app & 990 & 77 MB & 26192 & 630872 & 96647 & 402 & 1.40 & 2.44 & 42.65 \\
 & wideresnetapp & 1495 & 140 MB & 26352 & 631004 & 172712 & 303 & 20.02 & 55.38 & 63.85 \\
 & inceptionv3 & 4875 & 92 MB & 26862 & 1088344 & 250880 & 355 & 13.25 & 24.63 & 46.20 \\
 \bottomrule
\end{tabular}
}
\caption{\codename statistics for benchmarks and real-world applications.
Columns $2-3$: total application \loc and binary size.
Columns $4-7$: total \ocalls, system calls,
pagefaults, and context switches during one run.
Columns $8-9$: execution time on vanilla \dr and \tool.
Column $10$: \tool's execution overhead of \tool w.r.t. \dr. 
\tool performs better than \dr in some cases because of eager binary loading which improves cache hits.}
\label{tab:stats}
\end{table*} 
\paragraph{Library vs Binary Compatibility.}
We maintain full binary compatibility with all $203$ binaries tested
for which we had system call support in \codename. For them, \codename
works out-of-the-box in our experiments. We report that, given the same 
inputs as native execution, \codename produces same outputs.
With the aim of true binary compatibility, \codename supports binaries
without limiting them to a specific implementation or version of \libc.
To empirically demonstrate that our binary compatibility is superior,
we test \codename with binaries that use different \libc
implementations. Specifically, we compile HBenchOS benchmark ($12$
binaries) with \glibc v$2.23$ and \musl ~\libc v$1.2.0$.
We report that \codename executes these system stress workloads
out-of-the-box with both the libraries. We do not make any change to
our implementation. Lastly, we report our experience on porting our
micro-benchmarks to a state-of-the-art library-OS called \gsgx in
Appendix~\ref{appx:gsgx}. Of the $77$ programs tested, \gsgx fails on
$13$. \tool works out-of-the-box for all except $1$, which failed due
to the virtual memory limit enforced by SGX hardware.

\begin{table}[tbp]
\centering
\resizebox{0.5\textwidth}{!}{
\begin{tabular}{@{}clSSS@{}}
\toprule
\multirow{2}{*}{\textbf{Property}} & \multicolumn{1}{c}{\multirow{2}{*}{\textbf{Sub-property}}} & \multicolumn{2}{c}{\textbf{Performance}} \\ \cmidrule(l){3-4} 
 & \multicolumn{1}{c}{} & \multicolumn{1}{c}{\textbf{DR}} & \multicolumn{1}{c}{\textbf{Ratel}} \\ \midrule
\multirow{7}{*}{\begin{tabular}[c]{@{}c@{}}Memory \\ Intensive\\ Operations\\ Bandwidth (MB/s)\\ More iterations\\ Less Chunk \\size\end{tabular}} 
& Raw Memory Read & 24976.73 & 24665.05\\
 & Raw Memory Write & 12615.13 & 12580.36\\
& Bzero Bandwidth &  60877.42 &  65072.41\\
& Memory copy libc aligned & 56883.67 & 60377.04\\
& Memory copy libc unaligned & 56270.52 & 61543.81\\
& Memory copy unrolled aligned &  12272.93 &  12351.22\\
& Memory copy unrolled unaligned & 12279.55  & 12295.40\\
& Mmapped Read & 423.85 & 190.73 \\
& File Read & 29.15 & 12.05 \\
\midrule
\multirow{9}{*}{\begin{tabular}[c]{@{}c@{}}Memory \\ Intensive\\ Operations \\ Bandwidth (MB/s)\\ Less iteration \\ More Chunk \\size\end{tabular}} 
    & Raw Memory Read & 13292.24 & 5717.96\\
    & Raw Memory Write & 10664.76 & 4563.60 \\
    & Bzero Bandwidth &  31315.64 &  4166.62\\
    & Memory copy libc aligned & 12969.82 & 1570.94\\
    & Memory copy libc unaligned & 13141.00 & 1556.58\\
    & Memory copy unrolled aligned & 6714.93 & 2054.55\\
    & Memory copy unrolled unaligned & 5853.26  & 2081.05\\
    & Mmapped Read & 7163.38 & 3299.77 \\
    & File Read & 3724.39 & 769.66 \\ 
\midrule
\multirow{5}{*}{\begin{tabular}[c]{@{}c@{}}File \\System \\Latency(s) \end{tabular}} 
    & Filesystem create & 32.43 & 115.34 \\ 
    & Filesystem delforward & 18.41 & 33.53 \\
    & Filesystem delrand & 21.17 & 37.28 \\
    & Filesystem delreverse & 18.31 & 33.13 \\
    \midrule
\multirow{6}{*}{\begin{tabular}[c]{@{}c@{}}System\\ Call\\ Latency(s)\end{tabular}} 
 & getpid & 0.0065 & 0.0058 \\
 & getrusage & 0.64 & 7.45 \\
 & gettimeofday & 0.0239 & 6.5986 \\
 & sbrk & 0.0064 & 0.0065 \\
 & sigaction & 2.21  & 2.79 \\
 & write & 0.51 & 7.23 \\ \midrule
\multirow{2}{*}{\begin{tabular}[c]{@{}c@{}}Signal\\ Handler Latency\end{tabular}} 
 & Installing Signal & 2.24 & 2.79 \\
 & Handling Signal & 8.88 & 81.58 \\ 
\bottomrule
\end{tabular}
}
\caption{Summary of HBenchOS benchmark results.}
\label{tab:hbenchos}
\end{table} \begin{figure*}[tbp]
\centering
\begin{minipage}[h]{0.25\textwidth}
\includegraphics[width=0.9\textwidth,height=3cm]{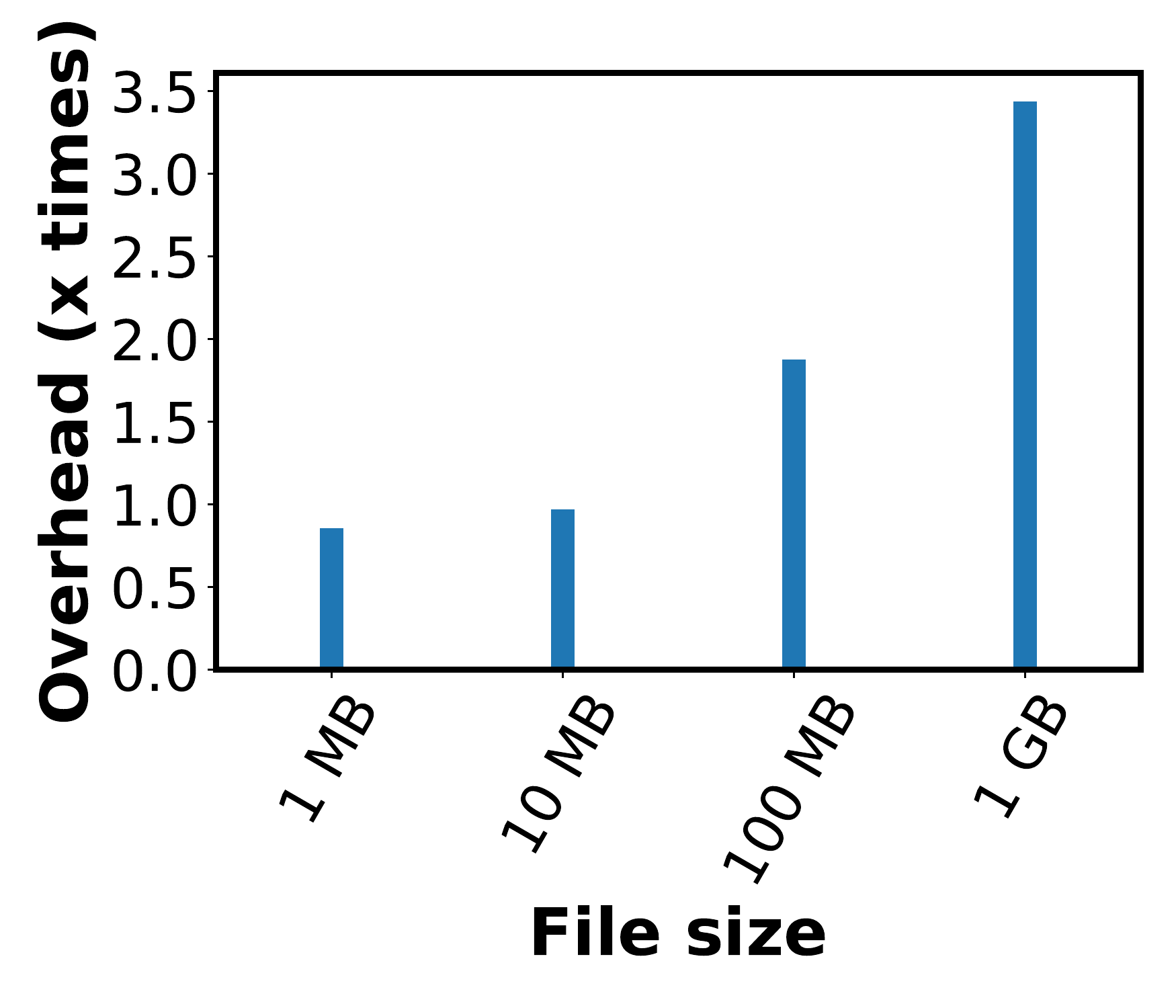}
\subcaption{cURL}	
\label{fig:curl}
\end{minipage}
\hspace{-5mm}
\begin{minipage}[h]{0.25\textwidth}
\includegraphics[width=\textwidth,height=3cm]{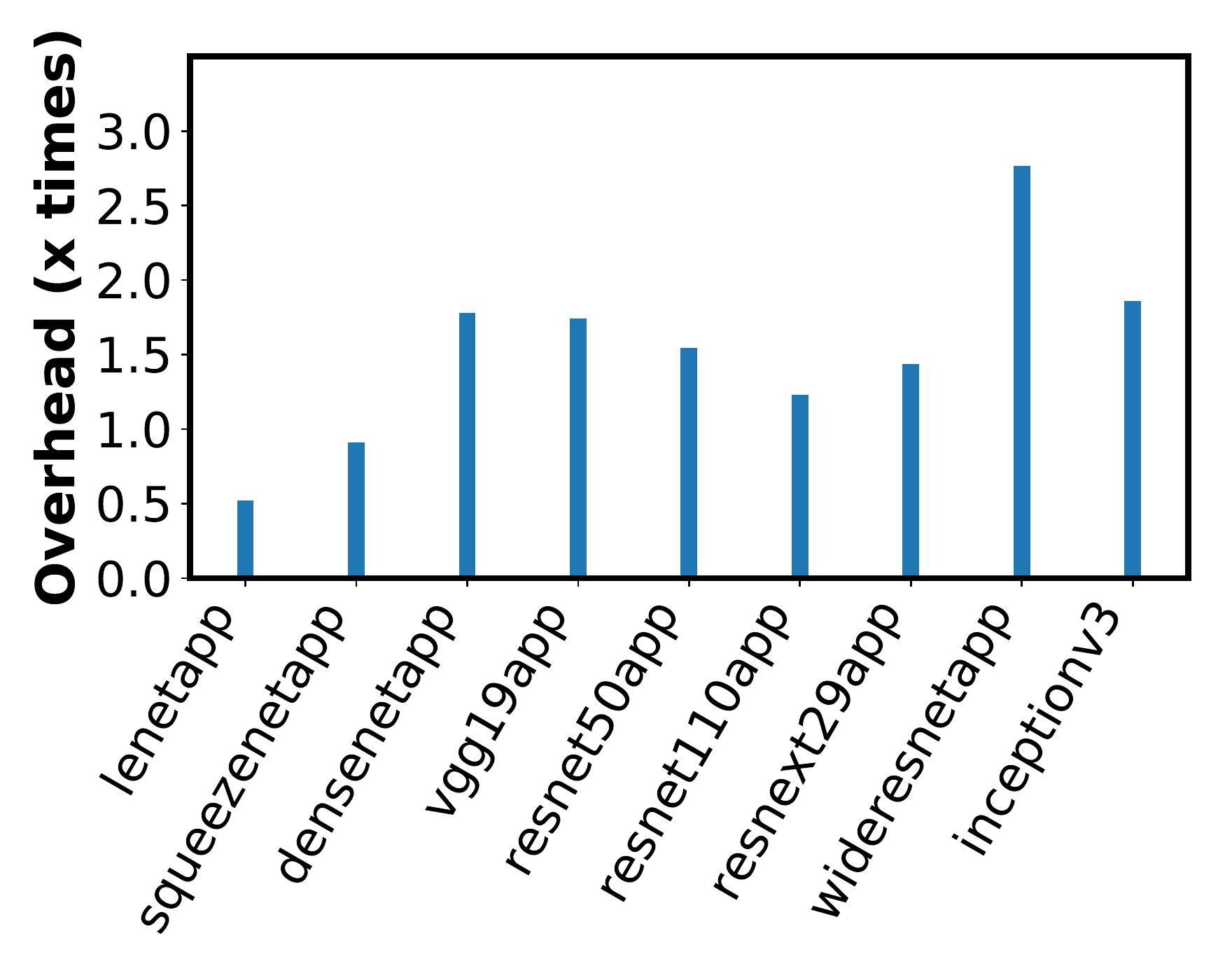}
\subcaption{Privado}	
\label{fig:privado}
\end{minipage}
\begin{minipage}[h]{0.25\textwidth}
    \includegraphics[width=\textwidth,height=3cm]{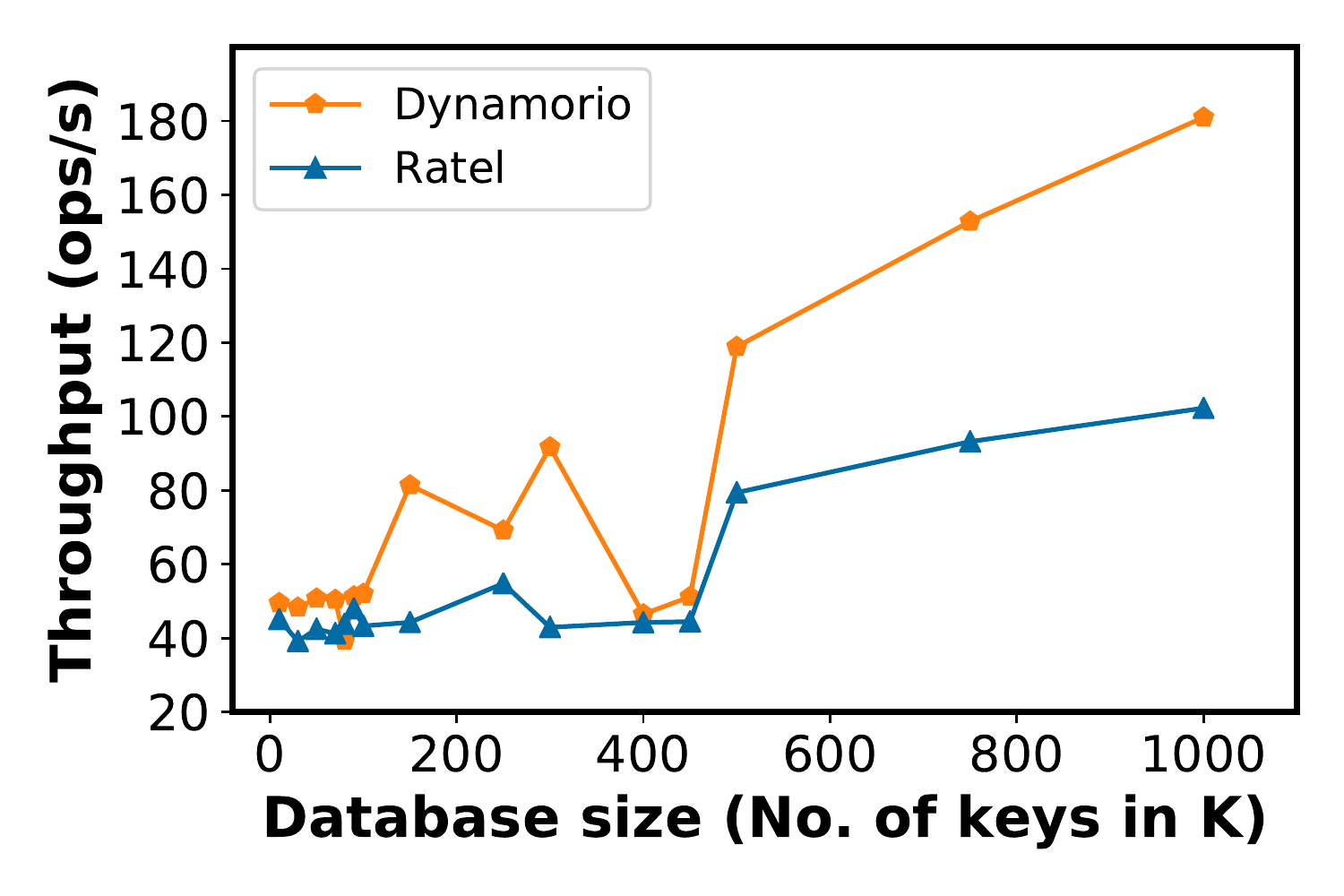}
    \subcaption{SQLite}	
    \label{fig:sqlite}
    \end{minipage}
    \begin{minipage}[h]{0.25\textwidth}
    \includegraphics[width=\textwidth,height=3cm]{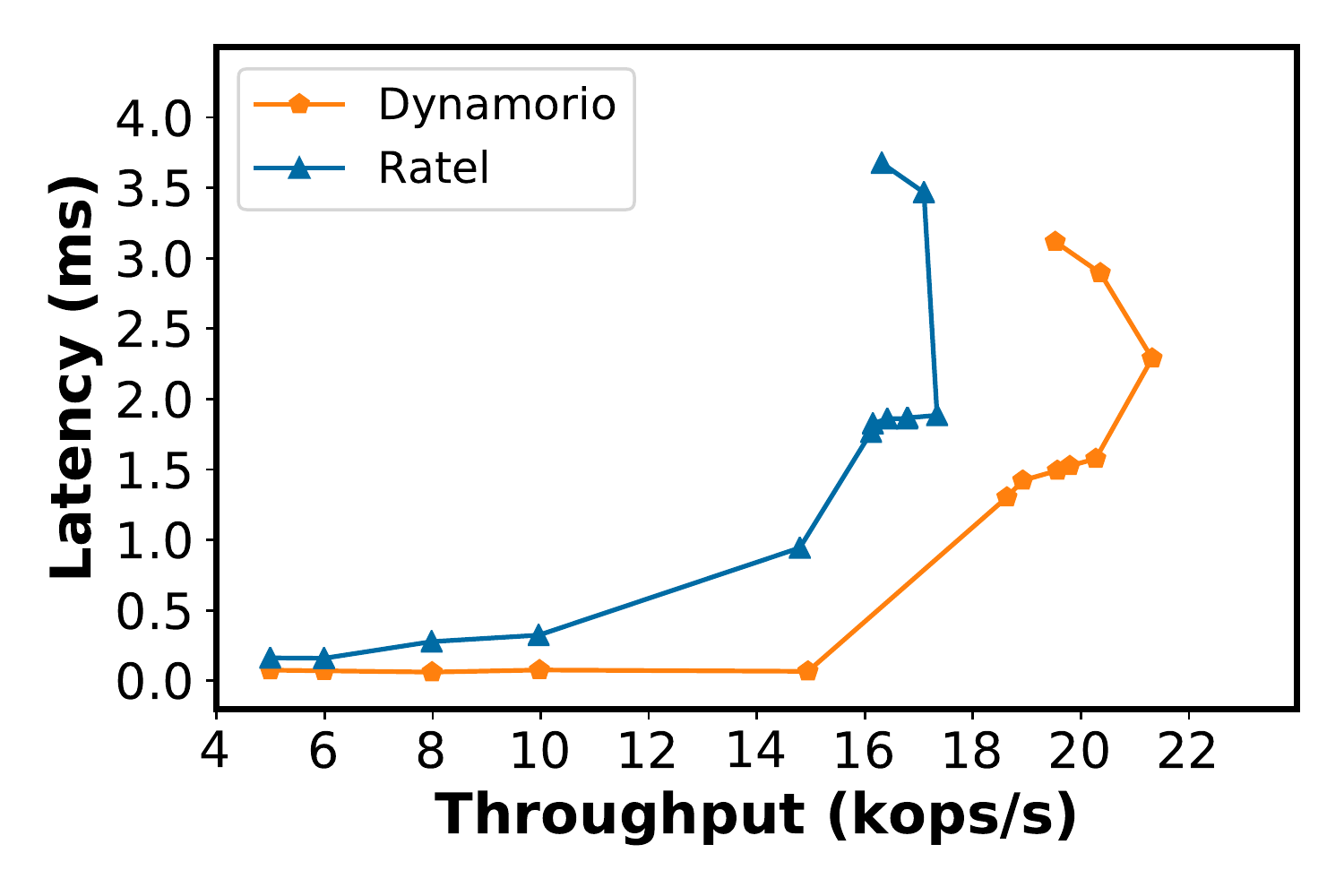}
    \subcaption{Memcached}	
    \label{fig:memcached}
    \end{minipage}
\vspace{-5pt}
\caption{\tool Performance.
(a) cURL,
(b) Privado,
(c) SQLite, and
(d) Memcached. 
(a) and (b) show \tool execution time overhead w.r.t. vanilla \dr. 
(c) shows average time per operation (micros/op) with increasing
database size represented as number of primary keys in thousands (K);
(d) shows the throughput versus latency.
}
\label{fig:apps-curl-privado}
\vspace{-15pt}
\end{figure*}

 \begin{figure*}[tbp]
\centering
\begin{minipage}[h]{0.32\textwidth}
    \includegraphics[width=\textwidth,height=4cm]{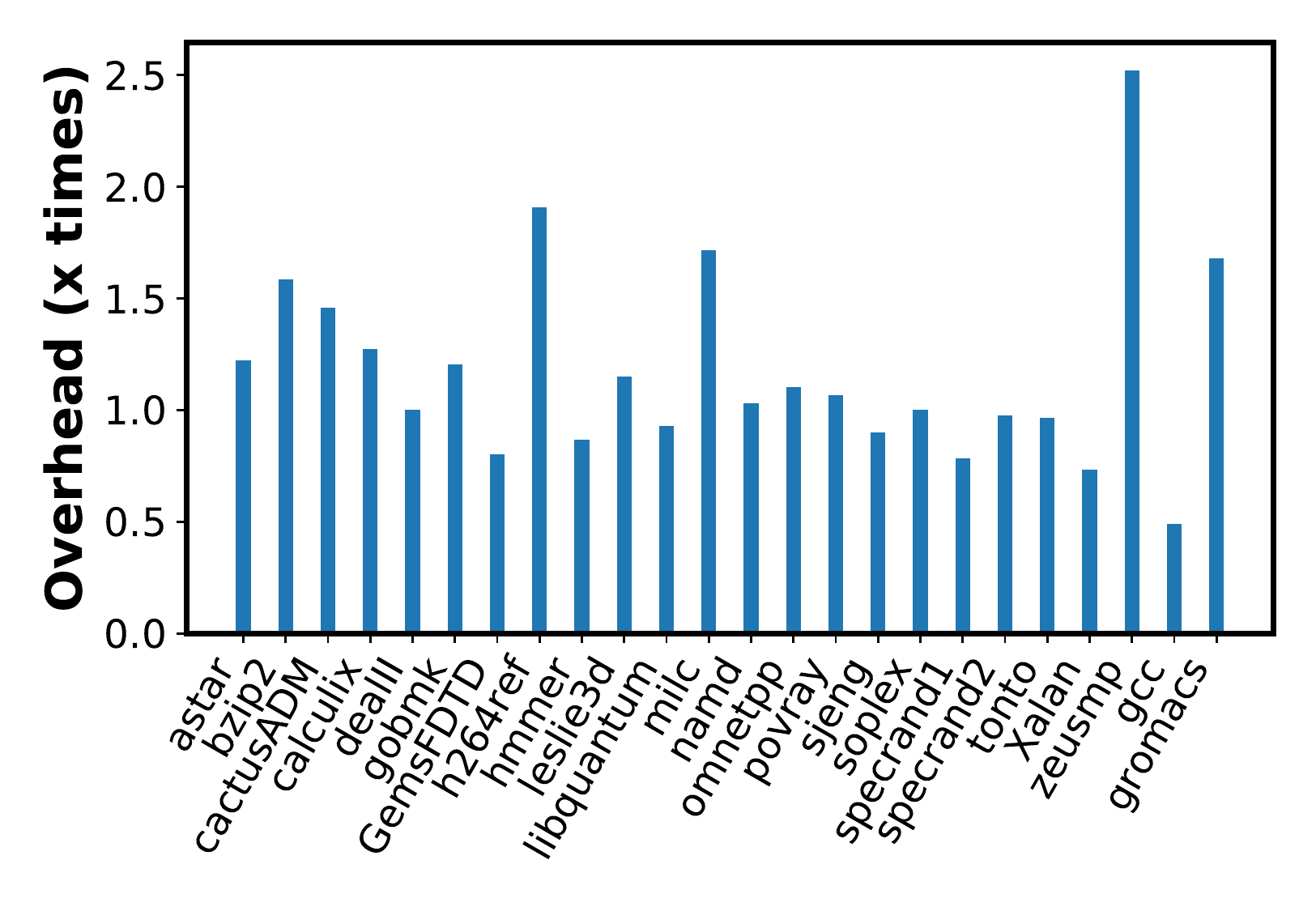}
    \subcaption{SPEC 2006}	
    \label{fig:spec}
    \end{minipage}
    \begin{minipage}[h]{0.32\textwidth}
    \includegraphics[width=\textwidth,height=4cm]{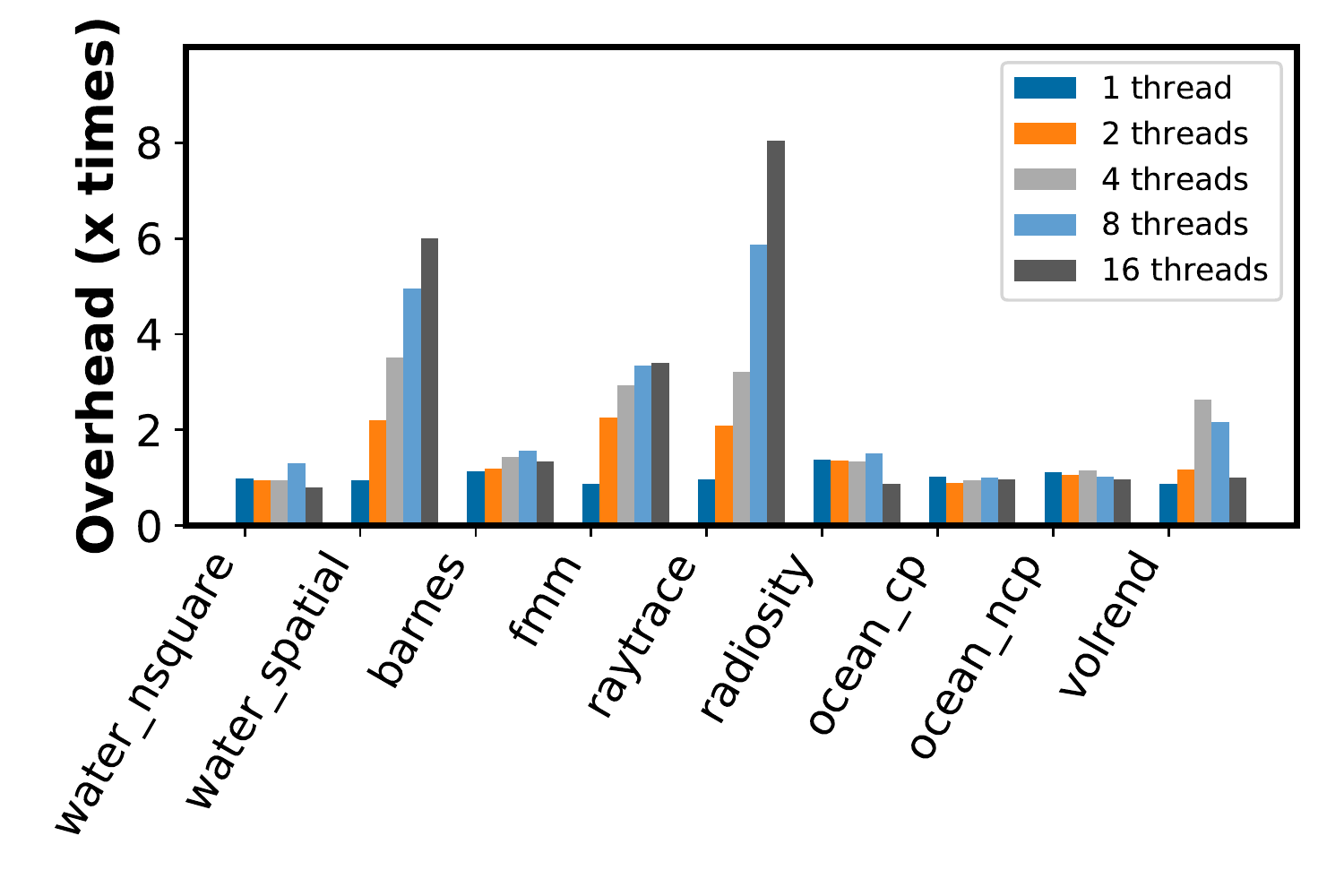}
    \subcaption{Parsec-SPLASH-2}	
    \label{fig:parsec}
    \end{minipage}
\begin{minipage}[h]{0.32\textwidth}
\includegraphics[width=\textwidth,height=4cm]{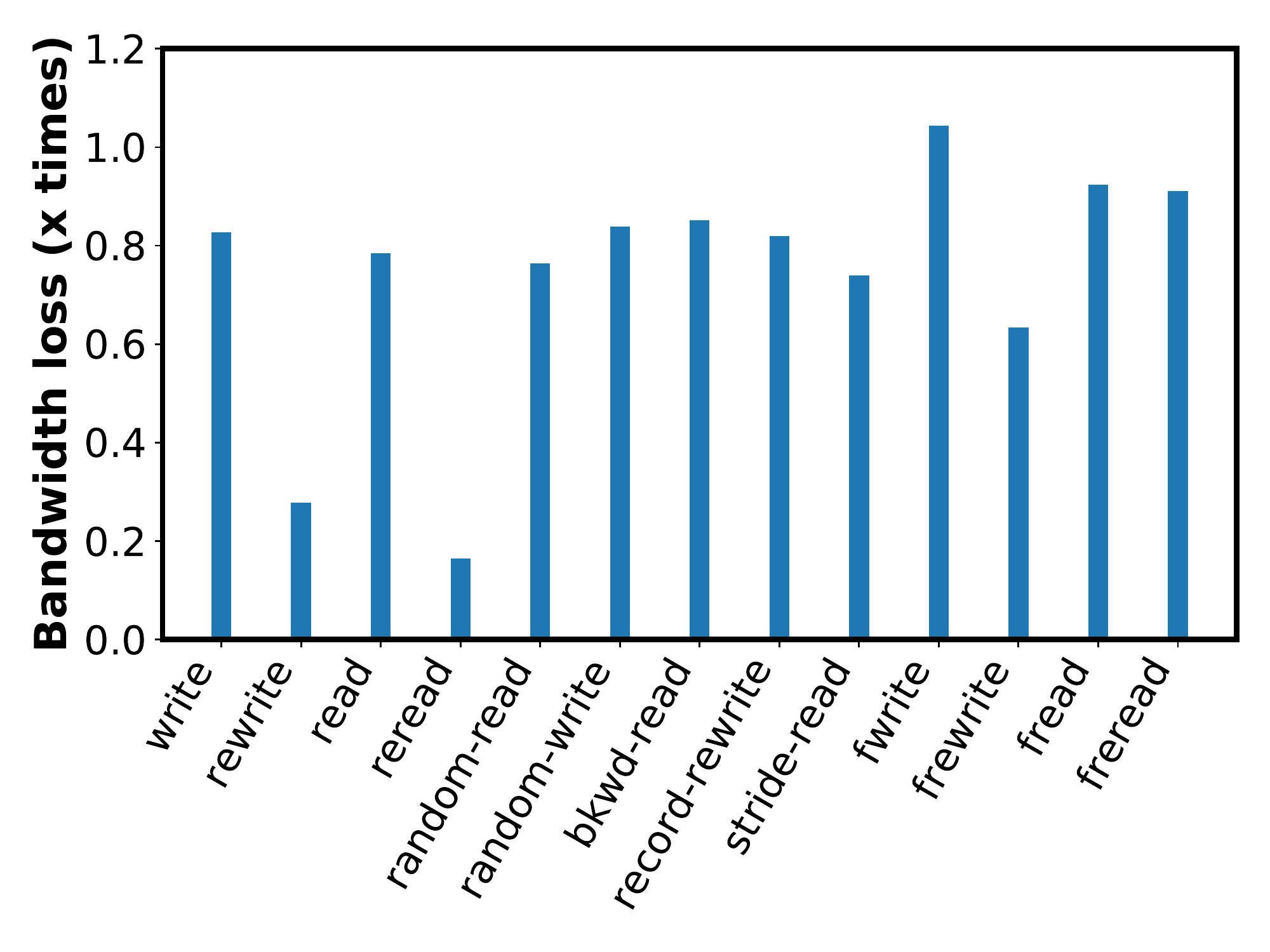}
\subcaption{IOZone}	
\label{fig:iozone}
\end{minipage}
\vspace{-5pt}
\caption{\codename Performance for micro-benchmarks:
(a) SPEC 2006 (CPU),
(b) Parsec-SPLASH-2 (multi-threading), and
(c) IOZone (I/O). 
(a) and (b) show \tool execution time overhead w.r.t. vanilla \dr;
lower value indicates better performance. (c) shows bandwidth loss
w.r.t. vanilla \dr; value close to $0$ indicates worse performance.}
\label{fig:benchmarks}
\vspace{-15pt}
\end{figure*}  \subsection{TCB Breakdown}
\label{sec:tcb}

Since we aim to run applications inside enclaves, we trust Intel SGX
support software (SDK, PSW) that allows us to interface with the
hardware. This choice is same as any other system that uses enclaves.
\codename comprises one additional trusted components i.e., \dr. Put
together, \codename amounts to $277,803$ \loc TCB. This is
comparable to existing SGX frameworks that have $100$K to $1$M
\loc~\cite{scone, graphene-sgx}, but only provide library-based
compatibility.

Table~\ref{tab:tcb} (columns $2$-$4$) summarizes the breakdown of the
\loc included in the trusted components of the PSW, SDK, and \dr as
well as the code contributed by each of the sub-systems supported by
\codename.
Original \dr comprises $353,139$ \loc. We reduce it to $129,875$ \loc
(trusted) and $66,629$ \loc (untrusted) by removing the components
that are not required or used by \codename. Then we add $8,589$ \loc
to adapt \dr to SGX as per the design outlined in
Section~\ref{sec:design}. Apart from this, as described in
Section~\ref{sec:impl}, we change the libraries provided by Intel SGX
(SDK, PSW, Linux driver) and add $1,078$ \loc. 

Of the $277,803$ \loc of trusted code, $123,322$ \loc is from the
original \dr code base responsible for loading the binaries,
code-cache management, and syscall handling. $110,848$ \loc and
$37,080$ \loc are from Intel SGX SDK and PSW respectively. 
\codename implementation adds only $6,553$ \loc on top of this
implementation. 
A large fraction of our added TCB ($27.5$\%) is because of the \ocall
wrappers that are amenable to automated testing and
verification~\cite{coins, besfs}. Rest of the $4,752$ \loc are for
memory management, handling signals, TLS, and multi-threading
interface. 

\codename relies on, but does not trust, the code executing outside
the enclave in the host process (e.g., \ocalls). This includes $2,391$
\loc changes, Table~\ref{tab:tcb} (columns $6$-$10$). \subsection{Performance Analysis}
\label{sec:perf-analysis}

\tool explicitly trades-off performance for secure and complete
mediation, by design. We present the performance implications
of these design choices.
We have two main findings. First, the performance overheads vary
significantly based on the application workload. Second, we find that
most of the overheads come stem from the specific SGX restrictions
R1-R5 and due to limited physical memory available.
 
To measure these, we collect various statistics of the execution
profile of $58$ program in our micro-benchmarks and $4$ real-world
applications ($12$ binaries in total). Specifically, we log the target
application \loc, binary size, number of \ocalls, \ecalls, syscalls,
enclave memory footprint (stack and heap), number of page faults, and
number of context switches. Table~\ref{tab:stats} shows these
statistics for the benchmarks and applications evaluated with
\codename. Interested readers can refer to Appendix~\ref{sec:appx-bm},
~\ref{sec:appx-apps} for detailed performance breakdowns.
There are three main avenues of overhead costs. 

First, fundamental limitations of SGX result in increased
memory-to-memory operations (e.g., two-copy design) or usage of slower
constructs (e.g., spin-locks instead of fast futexes). 
Our evaluation on system stress workloads for each subsystem measure
the worst-case cost of these operations.
We report that on an average, SPEC CPU benchmarks result in $40.24\%$
overheads (Figure~\ref{fig:spec}), while I/O-intensive workloads cost
$75\%$ slowdown (Figure~\ref{fig:iozone} for IOZone benchmarks).
Further, the performance overheads increases with larger I/O record
sizes. The same is observed for HBenchOS binaries as reported in
Table~\ref{tab:hbenchos}.
The expensive spin-locks incur cost that increases with number of
threads (Figure~\ref{fig:parsec} for Parsec-SPLASH2 benchmarks).
Overall, we observe that benchmarks that require large memory copies 
consistently exhibit significant slow-downs compared to others,
highlighting the costs imposed by the two-copy design. The cost of
signal handling also increases due to added context saves and restores
in \codename, as seen in a dedicated benchmark of HBenchOS (see last
two rows in Table~\ref{tab:hbenchos}). 

We believe some of these observed costs will be common to other
compatibility engines, while the remaining stem from our preference
for binary compatibility in designing \tool. As an example,
our evaluation on \gsgx (Appendix~\ref{appx:gsgx}) shows
memory-intensive workloads exhibit a similar increase in overheads.
\gsgx does not use spin-locks and tunnels all signal handling through \libc 
as it prioritizes performance over binary compatibility,
and has reduced overheads compared to \tool.

Second, the current SGX hardware implementation has limited secure
physical memory (called the EPC), only $90$ MB. Executing anything on
a severely limited memory resource results in large slow-downs (e.g.,
increased page-faults). Further, cost of each page-in and page-out
operation itself is higher in SGX because of hardware based memory
encryption.
We measure the impact of this limitation by executing benchmarks and
applications that exceed the working set size of $90$ MB for both data
and code.
For example, we test varying download sizes in cURL
(Figure~\ref{fig:curl}) and database sizes in SQLite
(Figure~\ref{fig:sqlite}). When the data exceeds $90$ MB, we observe a
sharp increase in throughput loss.
Similarly, when we execute varying sizes of ML models that require
increasing size of code page memory, we observe increase in page
faults and lowered performance (Figure~\ref{fig:privado}).
We observe similar loss of latency and throughput when applications
reach a critical point in memory usage as in $\tt{memcached}$
(Figure~\ref{fig:memcached}). Detailed performance breakdown for these
applications is in Appendix~\ref{sec:appx-apps}.

The added overhead is solely because of \tool is the cost of dynamic
binary translation itself. In the original \dr, the DBT design
achieves close to native or better after the code-caches warm
up~\cite{dynamorio}. In \codename, we expect to preserve the same
performance characteristics as \dr. However, the SGX physical memory
limits directly impact the execution profile of \dr running in the
enclave. Specifically, \codename may result in an increase in
physical memory for its own execution that may slow-down the target
binary. It is difficult to directly measure the exact cost incurred
by this factor because we cannot increase the hardware
physical memory in our setup.

Lastly, we observe that the performance costs variation based on
workloads are common to other platforms. As a direct comparison point,
we tested HBenchOS---a benchmark with varying workloads---with
Graphene-SGX. Graphene-SGX is a popular and well-maintained library-OS
has been under active development for several years as of this
writing.
Interested readers can refer to Appendix~\ref{appx:gsgx}, ~\ref{sec:appx-apps} 
for details. \tool offers better binary compatibility as opposed to
Graphene-SGX which provides compatibility with \glibc.
  \section{Related Work}
\label{sec:related}

Several prior works have targeted SGX compatibility. There are two
main ways that prior work has overcome these challenges. The first
approach is to fix the application interface. The target application
is either re-compiled or is relinked to use such interfaces. The
approach that enables the best compatibility exposes specific Libc
(\texttt{glibc} or \texttt{musl} libc) versions as interfaces. This
allows them to adapt to SGX restrictions at a layer below the
application. Container or libraryOS solutions use this to execute
re-compiled/re-linked code inside the enclave as done in
Haven~\cite{haven}, Scone~\cite{scone}, \gsgx~\cite{graphene-sgx},
Ryoan~\cite{ryoan}, SGX-LKL~\cite{sgx-lkl}, and Occlum~\cite{occlum}.
Another line of work is compiler-based solutions. They require applications to
modify source code to use language-level
interface~\cite{panoply,baidu-rust-sgx, mesatee-sgx, fortanix-rust-sgx}.

Both style of approaches can have better performance than \tool, but
require recompiling or relinking applications. For example, library
OSes like \gsgx and containerization engines like Scone expose a
particular \glibc and \musl version that applications are asked to
link with. New library versions and interfaces can be ported
incrementally, but this creates a dependence on the underlying
platform interface provider, and incurs a porting effort for each
library version. Applications that use inline assembly or runtime code
generation also become incompatible as they make direct access to
system calls, without using the API. \tool approach of handle R1-R5
comprehensively offers secure and complete mediation, without any
assumptions about specific interfaces beyond that implied by binary
compatibility.

\paragraph{Security Considerations.}
As in \codename, other approaches to SGX compatibility eventually have
to use \ocalls, \ecalls, and syscalls to exchange information between
the enclave and the untrusted software. This interface is known to be
vulnerable~\cite{tale-two-worlds, iago}. Several shielding systems for
file~\cite{besfs, sgx-fs} and network IO~\cite{talos}, provide
specific mechanisms to safeguard the OS interface against these
attacks. For security, defense techniques offer compiler-based tools
for enclave code for memory safety~\cite{sgxbounds},
ASLR~\cite{sgx-shield}, preventing controlled-channel
leakage~\cite{tsgx}, data location randomization~\cite{drsgx}, secure
page fault handlers~\cite{cosmix}, and branch information
leakage~\cite{privado}.

\paragraph{Performance.}
Several other works build optimizations by modifying existing
enclave-compliant library OSes. One such example is
Hotcalls~\cite{hotcalls}, Eleos~\cite{eleos} which add exit-less calls
to reduce the overheads of \ocalls. These well-known optimizations are
also available as part of the default Intel SGX SDK now.

\paragraph{Language Run-times.}
Recent body of work has also shown how executing either
entire~\cite{sgx-lang-interpreter} or partial~\cite{civet} language
runtimes inside an enclave can help to port existing code written in
interpreted languages such as Python~\cite{python-sgx, mesa-py},
Java~\cite{civet}, web-assembly~\cite{acctee}, Go~\cite{go-tee}, and
JavaScript~\cite{trustjs}.

\paragraph{Programming TEE Applications.}
Intel provides a C/C++ SGX software stack which includes a SDK and OS
drivers for simulation and PSW for running local enclaves. There are
other SDKs developed in in memory safe languages such as
Rust~\cite{baidu-rust-sgx, mesatee-sgx, fortanix-rust-sgx}. Frameworks
such as Asylo~\cite{asylo}, OpenEnclave~\cite{open-enclave}, and
MesaTEE~\cite{mesatee} expose a high-level front-end for writing
native TEE applications using a common interface. They support several
back-end TEEs including Intel SGX and ARM TrustZone. Our experience
will help them to sidestep several design challenges as well as
improve their forward compatibility. 

\paragraph{Future TEEs.}
New enclave TEE designs have been
proposed~\cite{podarch,bastion,keystone,sanctum, komodo}.
Micro-architectural side channels~\cite{mi6} and new oblivious
execution capabilities~\cite{sanctum,phantom} are significant concerns
in these designs.
Closest to our underlying TEE is the recent Intel SGX v2~\cite{sgx2,
  sgxv2, sgx-sw2}. SGX v2 enables dynamically memory and thread
management inside the enclave, thus addressing $R2$ to some
extent. The other restrictions are largely not addressed in SGX v2,
and therefore, \tool design largely applies to it as well. Designing
enclave TEEs that do not place the same restrictions as SGX remain
promising future work.
 \section{Conclusion}
\label{sec:conclusion}

We present the design of \tool, the first work to offer {\em binary
  compatibility} with existing software on SGX. \tool is a dynamic
binary translation engine inside SGX enclaves on Linux. Through the
lens of \tool, we expose the fundamental trade-offs between performance
and ensure secure mediation on the OS-enclave interface. These
trade-offs are rooted in $5$ SGX design restrictions, which offer
concrete challenges to next-generation enclave TEE designs.
 \section*{Acknowledgments}
We thank David Kohlbrenner, Zhenkai Liang, and Roland Yap for their
feedback on improving earlier drafts of the paper. We thank Shipra
Shinde for help on formatting the figures in this paper.
This research was partially supported by a grant from the National
Research Foundation, Prime Ministers Office, Singapore under its
National Cybersecurity R\&D Program (TSUNAMi project, No.
NRF2014NCR-NCR001-21) and administered by the National Cybersecurity
R\&D Directorate. 
This material is in part based upon work supported by the National
Science Foundation under Grant No. DARPA N66001-15-C-4066 and Center
for Long-Term Cybersecurity. Any opinions, findings, and conclusions
or recommendations expressed in this material are those of the
authors and do not necessarily reflect the views of the National
Science Foundation. \section*{Availability}

\codename implementation, including modified Intel SGX SDK, PSW, and
driver, is available at \url{https://ratel-enclave.github.io/}
Our project webpage and GitHub repository also contains unit
tests, benchmarks, Linux utils, and case studies evaluated in this
paper. 
  
\balance

\bibliographystyle{plain}
\def\UrlBreaks{\do\/\do-}

\appendix

\subsection{Details of Compatibility Tests with \tool}
\label{sec:appx-utilities}

\begin{table*}[tbp]
\centering
\resizebox{\textwidth}{!}{
\begin{tabular}{@{}lc|lc|lc|lc|lc|lc@{}}
\toprule
\textbf{Utility} & \textbf{\# of syscalls} & \textbf{Utility} & \textbf{\# of syscalls} & \textbf{Utility} & \textbf{\# of syscalls} & \textbf{Utility} & \textbf{\# of syscalls} & \textbf{Utility} & \textbf{\# of syscalls} & \textbf{Utility}                   & \textbf{\# of syscalls} \\ \midrule
ed      & 18             & ppdpo   & 29             & dirmngr & 21             & hcitool    & 20             & systemctl     & 32             & systemd-cgtop                      & 25             \\
cvt     & 13             & psnup   & 14             & enchant & 20             & bluemoon   & 21             & vim.basic     & 45             & dirmngr-client                     & 13             \\
eqn     & 36             & t1asm   & 13             & epsffit & 13             & btattach   & 21             & hciconfig     & 25             & systemd-escape                     & 18             \\
gtf     & 18             & troff   & 15             & faillog & 18             & fwupdate   & 19             & brltty-ctb    & 21             & systemd-notify                     & 19             \\
pic     & 13             & uconv   & 13             & gendict & 14             & gatttool   & 24             & fusermount    & 19             & wpa\_passphrase                    & 18             \\
tbl     & 13             & bccmd   & 25             & hex2hcd & 18             & gencnval   & 13             & journalctl    & 26             & gamma4scanimage                    & 13             \\
xxd     & 14             & btmgmt  & 24             & icuinfo & 15             & lessecho   & 13             & sudoreplay    & 16             & systemd-analyze                    & 31             \\
curl    & 32             & busctl  & 37             & kbxutil & 14             & loginctl   & 31             & watchgnupg    & 18             & systemd-inhibit                    & 31             \\
derb    & 23             & catman  & 16             & lastlog & 14             & makeconv   & 14             & xmlcatalog    & 12             & systemd-resolve                    & 31             \\
find    & 27             & cd-it8  & 21             & lesskey & 13             & ppdmerge   & 26             & zlib-flate    & 13             & ulockmgr\_server                   & 18             \\
gawk    & 25             & expiry  & 14             & lexgrog & 15             & psresize   & 14             & cupstestdsc   & 26             & systemd-tmpfiles                   & 34             \\
grep    & 21             & genbrk  & 14             & manpath & 14             & psselect   & 14             & cupstestppd   & 18             & gpg-connect-agent                  & 22             \\
htop    & 26             & gencfu  & 13             & obexctl & 35             & t1binary   & 13             & hostnamectl   & 30             & kerneloops-submit                  & 20             \\
kmod    & 17             & grotty  & 13             & pkgdata & 14             & t1binary   & 13             & systemd-run   & 29             & evince-thumbnailer                 & 21             \\
ppdc    & 30             & l2ping  & 21             & ppdhtml & 27             & t1disasm   & 13             & timedatectl   & 32             & fcitx-dbus-watcher                 & 18             \\
ppdi    & 30             & l2test  & 27             & preconv & 15             & t1disasm   & 13             & brltty-trtxt  & 22             & systemd-detect-virt                & 18             \\
qpdf    & 14             & psbook  & 14             & sdptool & 22             & transfig   & 15             & dbus-monitor  & 31             & dbus-cleanup-sockets               & 19             \\
gpg2    & 27             & pstops  & 14             & ssh-add & 20             & vim.tiny   & 34             & dbus-uuidgen  & 13             & systemd-stdio-bridge               & 25             \\
wget    & 29             & rctest  & 25             & t1ascii & 13             & dbus-send  & 30             & fcitx-remote  & 23             & systemd-ask-password               & 20             \\
btmon   & 23             & soelim  & 12             & udevadm & 27             & gpg-agent  & 18             & gpgparsemail  & 14             & webapp-container-hook              & 26             \\
genrb   & 13             & whatis  & 20             & volname & 15             & hciattach  & 18             & systemd-hwdb  & 17             & systemd-machine-id-setup           & 18             \\
grops   & 14             & rfcomm  & 19             & xmllint & 15             & localectl  & 30             & systemd-path  & 18             & systemd-tty-ask-password-agent     & 25             \\
mandb   & 27             & bootctl & 19             & ciptool & 20             & pg\_config & 16             & enchant-lsmod & 13             & dbus-update-activation-environment & 31                          \\ \bottomrule
\end{tabular}
}
\caption{List of GNU utilities ($138$) tested with \codename along with the number of unique system calls called during a single run.}
\label{tab:utilities}
\end{table*} \begin{table*}[tbp]
\centering
\resizebox{\textwidth}{!}{
\begin{tabular}{@{}lc|lc|lc|lc|lc|lc@{}}
\toprule
\textbf{Utility} & \textbf{\# of syscalls} & \textbf{Utility} & \textbf{\# of syscalls} & \textbf{Utility} & \textbf{\# of syscalls} & \textbf{Utility} & \textbf{\# of syscalls} & \textbf{Utility} & \textbf{\# of syscalls} & \textbf{Utility}           & \textbf{\# of syscalls} \\ \midrule
gcc              & 43                      & dealII           & 43                      & leslie3d         & 43                      & xalancbmk        & 49                      & bw\_mmap\_rd     & 42                      & water\_spatial             & 44                      \\
fmm              & 45                      & soplex           & 43                      & calculix         & 41                      & LFS-write        & 43                      & resnet50app      & 41                      & inceptionv3app             & 42                      \\
curl             & 55                      & povray           & 43                      & GemsFDTD         & 42                      & multiopen        & 41                      & densenetapp      & 42                      & multicreatemany            & 41                      \\
milc             & 42                      & barnes           & 45                      & specrand         & 42                      & multiread        & 41                      & multicreate      & 47                      & multicreatewrite           & 41                      \\
namd             & 42                      & iozone           & 47                      & specrand         & 42                      & memcached        & 59                      & lat\_fslayer     & 42                      & lat\_syscall(sbrk)         & 42                      \\
bzip2            & 48                      & lat\_fs          & 41                      & lenetapp         & 41                      & radiosity        & 44                      & lat\_connect     & 42                      & lat\_syscall(write)        & 42                      \\
gobmk            & 42                      & bw\_tcp          & 42                      & vgg19app         & 43                      & ocean\_ncp       & 44                      & resnext29app     & 42                      & lat\_syscall(getpid)       & 42                      \\
hmmer            & 41                      & h264ref          & 42                      & raytrace         & 45                      & bw\_mem\_cp      & 42                      & resnet110app     & 41                      & lat\_syscall(sigaction)    & 42                      \\
sjeng            & 42                      & omnetpp          & 43                      & ocean\_cp        & 45                      & bw\_mem\_rd      & 42                      & wideresnetapp    & 43                      & lat\_syscall(getrusage)    & 43                      \\
tonto            & 44                      & gromacs          & 46                      & volerand         & 45                      & bw\_mem\_wr      & 42                      & squeezenetapp    & 41                      & lat\_syscall(gettimeofday) & 42                      \\
astar            & 43                      & lat\_sig         & 44                      & bw\_bzero        & 42                      & libquantum       & 42                      & LFS-smallfile    & 49                      &                            & \multicolumn{1}{l}{}    \\
sqlite           & 47                      & lat\_tcp         & 42                      & lat\_mmap        & 42                      & multiwrite       & 41                      & LFS-largefile    & 42                      &                            & \multicolumn{1}{l}{}    \\
zeusmp           & 43                      & lat\_udp         & 42                      & cactusADM        & 43                      & bw\_file\_rd     & 42                      & water\_nsquare   & 44                      &                            & \multicolumn{1}{l}{}    \\ \bottomrule
\end{tabular}
}
\caption{List of applications ($12$) and individual benchmarks ($63$) tested with \codename along with the number of unique system calls called during a single run.}
\label{tab:bapp}
\end{table*} \begin{table}[tbp]
\centering
\resizebox{0.5\textwidth}{!}{
\begin{tabular}{lcl}
\toprule 
\textbf{Reason category} & \textbf{\# of the unsuccessful} & \textbf{Case examples}                                                                           \\ \midrule
fork &	49 &	strace, scp, lat\_proc and lat\_pipe from HBenchOS, etc.
\\
execv &	1 &	systemd-cat
\\
signal &	5 &	colormgr, cd-iccdump, bluetoothctl etc.
\\
Unsupported syscalls &	6 &	webapp-container, webbrowser-app, etc.
\\
Out-of-memory &	3 &	shotwell, mcf from SPEC2006, lat\_memsize from HBenchOS \\
\bottomrule
\end{tabular}
}
\caption{Summary of the reasons for failure of all $64$ unsuccessful binaries tested with \codename.}
\label{tab:ursum}
\end{table} \begin{table}[tbp]
\centering
\resizebox{0.5\textwidth}{!}{
\begin{tabular}{lcl}
\toprule 
\textbf{Reason category} & \textbf{\# of the unsuccessful} & \textbf{Case examples}                                                                           \\ \midrule
NTFS related &	16 &	ntfs-3g, ntfs-3g.probe, ntfs-3g.secaudit, etc.
\\
Printer related &	7 &	lp, lpoptions, lpq, lpr, lprm, etc.
\\
Scanner related &	2 &	sane-find-scanner, scanimage
\\
Failure in native run &	5 &	umax\_pp, cd-create-profile, and bwaves from SPEC2006, etc.
\\
Failure in Dynamorio run &	8 &	ssh, ssh-keygen, dig, etc. \\
\bottomrule
\end{tabular}
}
\caption{Summary of the reasons for failure of all $38$ unsuccessful binaries tested with native and \dr.}
\label{tab:undrsum}
\end{table} 
\paragraph{Linux Utilities.}
We test the compatibility offered by \ratel with all the Linux
built-in binaries on our experimental Ubuntu system. These comprise
$229$ shared-objected binaries in total, which are typically are in
the directories $\tt{/bin}$ and $\tt{/usr/bin}$. 

We run each utility with the most representative options and inputs.
Out of $229$, our test machine natively and with \dr work with $195$.
Of these $195$ binaries, a total of $138$ have all system calls
presently supported in \tool, all of which worked correctly in our
tests out-of-the-box. The $57$ programs that did not work fail for $2$
reasons: missing syscall support and virtual memory limits imposed by
SGX. Table~\ref{tab:utilities} and Table~\ref{tab:bapp} list all Linux
utilities and binaries from real-world applications and benchmarks
that ran successfully, and present the number of unique system calls
for each. Table~\ref{tab:ursum} and Table~\ref{tab:undrsum} summarize
the reasons for all binaries that fail in \codename and in native and
\dr, respectively.

$45$ of the failing utilities are due to lack of multi-processing
($\tt{fork}$) support in \codename. $5$ utilities use
certain POSIX signals for which we have not supported completely (e.g., real-time signals $\tt{SIGRTMIN+n}$.). $6$
utilities fail because they invoke other unsupported system calls in
\codename, most of which the restriction $R3$ in SGX fundamentally
does not permit (e.g., shared memory syscalls such as $\tt{shmat}$,
$\tt{shmdt}$, $\tt{shmctl}$, etc.). $1$ utility fails because of
the virtual memory limit in SGX, as it loads more than $100$ shared
libraries. It should be noted that the $\tt{ioctl}$
syscall involves more than $100$ variable parameters, \codename
syscall stubs currently does not cover all of them.

\paragraph{Other Benchmarks \& Applications.} 
From the $81$ binaries from micro-benchmarks and real applications,
$7$ do not work with \codename. $5$ binaries from HBenchOS
($\tt{lat\_proc}$, $\tt{lat\_pipe}$, $\tt{lat\_ctx}$,
$\tt{lat\_ctx2}$, $\tt{bw\_pipe}$) either use \texttt{fork}
or shared memory system calls disallowed by $R3$.
$2$ binaries ($\tt{lat\_memsize}$ from HBenchOS, mcf from SPEC2006)
with \dr require virtual memory larger than SGX limits
on our experimental setup.
 \subsection{Comparison to \gsgx}
\label{appx:gsgx}

To compare \codename's binary compatibility and performance with other
approaches, we have chosen \gsgx, a library-OS which runs inside SGX
enclave. \gsgx offers the lowest compatibility barrier of all prior
systems to our knowledge, specifically offering compatibility with
$\tt{glibc}$.

\paragraph{Compatibility.} 
Applications using \gsgx have to work only with a specific library
interface, namely a custom $\tt{glibc}$, which requires re-linking
and build process changes. \tool, in contrast, has been designed for
binary compatibility which is a fundamental difference in design. To
demonstrate the practical difference, we reported in
Section~\ref{sec:compat} that HBenchOS benchmark works out-of-the-box
when if built with both $\tt{glibc}$ and $\tt{musl}$, as an example. 

\gsgx requires a manifest file for each application, that specifies
the main binary name as well as dynamic-libraries, directories, and
files used by the application. By default \gsgx does not allow
creation of new files during runtime. We use the
$\tt{allow\_file\_creation}$ to disable this default. We tested all
$77$ benchmark and application binaries (HBench-OS, Parsec-SPLASH2,
SPEC, IOZone, FSCQ, SQLite, cURL, Memcached, Privado), out of which
$64$ work with \gsgx. {\em Of the $13$ that fail on \gsgx, all except
$1$ work on \codename, with the only failure being due to virtual
memory limits.}

For \gsgx, $3/9$ Parsec-SPLASH2 binaries ($\tt{water\_nsquare}$,
$\tt{water\_spatial}$ and $\tt{volrend}$), IOZone binary, and SQLite
database workload \cite{leveldb} failed due to I/O error, (e.g.,
~\cite{filemaperr}) which is an open issue. $3/25$ binaries from
SPEC2006 failed. \gsgx fails for $\tt{cactusAMD}$ due failed due to a
signal failure, which is mentioned as an existing open issue on its
public project page~\cite{aexerr}. The $\tt{calculix}$ program fails
with a segmentation fault. The $\tt{omnetpp}$ could not process the
input file in-spite of making the input file as allowed in the
corresponding manifest file. $4$ networking related binaries from
HBench-OS namely $\tt{lat\_connect}$, $\tt{lat\_tcp}$, $\tt{lat\_udp}$
and $\tt{bw\_tcp}$ could not run, resulting in a ``bad address'' error
while connecting to \texttt{localhost}. $\tt{lat\_memsize}$ from
HBench-OaS fails on \gsgx as it fails on \tool too due to the virtual
memory limit.

\paragraph{Performance.}
We report the performance overheads of \gsgx for HBenchOS benchmarks
in Table~\ref{tab:hbenchos-graphene}, as compared to \dr baseline and
\tool. The slowdown in both the systems is comparable for I/O
benchmarks, since both of them incur two copies.
\gsgx is significantly faster than \dr baseline and \tool for system
call and signal handling, because it implements a library OS inside
the enclave and avoids expensive context switches. \tool delegates
most of the system calls to the OS and does not emulate it like \gsgx,
offering compatibility with multiple libraries in contrast. Further,
\ratel offers instruction-level instrumentation capability.

\begin{table}[tbp]
    \centering
    \resizebox{0.45\textwidth}{!}{
    \begin{tabular}{@{}clSSSS@{}}
    \toprule
    \multirow{2}{*}{\textbf{Property}} & \multicolumn{1}{c}{\multirow{2}{*}{\textbf{Sub-property}}} & \multicolumn{3}{c}{\textbf{Performance}} \\ \cmidrule(l){3-5} 
     & \multicolumn{1}{c}{}  & \multicolumn{1}{c}{\textbf{DR}} & \multicolumn{1}{c}{\textbf{Ratel}} & \multicolumn{1}{c}{\textbf{Graphene}} \\ \midrule
    \multirow{9}{*}{\begin{tabular}[c]{@{}c@{}}Memory \\ Intensive\\ Operations\\ Bandwidth (MB/s)\\ More iteration\\ Less Chunk \\size\end{tabular}} 
        & Raw Memory Read & 24976.73 & 24665.05 & 23210.89\\
        & Raw Memory Write & 12615.13 & 12580.36 & 12114.76 \\
        & Bzero Bandwidth &  60877.42 &  65072.41 &  47844.32\\
        & Memory copy libc aligned & 56883.67 & 60377.04 & 63423.44\\
        & Memory copy libc unaligned & 56270.52 & 61543.81 & 69444.44\\
        & Memory copy unrolled aligned &  12272.93 &  12351.22 & 12161.04\\
        & Memory copy unrolled unaligned & 12279.55  & 12295.40 & 10079.86\\
        & Mmapped Read & 423.85 & 190.73 & 3814.69\\
        & File Read & 29.15 & 12.05 & 325.52\\ 
    \midrule
    \multirow{9}{*}{\begin{tabular}[c]{@{}c@{}}Memory \\ Intensive\\ Operations \\ Bandwidth (MB/s)\\ Less iteration \\ More Chunk \\size\end{tabular}} 
    & Raw Memory Read & 13292.24 & 5717.96 & 5310.06\\
    & Raw Memory Write & 10664.76 & 4563.60 & 3495.86\\
    & Bzero Bandwidth &  31315.64 &  4166.62 &  4046.79\\
    & Memory copy libc aligned & 12969.82 & 1570.94 & 1534.59\\
    & Memory copy libc unaligned & 13141.00 & 1556.58 & 1545.08\\
    & Memory copy unrolled aligned & 6714.93 & 2054.55 &  2009.46\\
    & Memory copy unrolled unaligned & 5853.26  & 2081.05 & 1999.58\\
    & Mmapped Read & 7163.38 & 3299.77 & 1454.30\\
    & File Read & 3724.39 & 769.66 & 134.42\\ 
    \midrule
    \multirow{5}{*}{\begin{tabular}[c]{@{}c@{}}File \\System \\Latency(s) \end{tabular}} 
        & Filesystem create & 32.43 & 115.34 &  1272.86 \\ 
        & Filesystem delforward & 18.41 & 33.53 &  1185.10 \\
        & Filesystem delrand & 21.17 & 37.28 &  1073.69 \\
        & Filesystem delreverse & 18.31 & 33.13 &  1266.38 \\
        \midrule
    \multirow{6}{*}{\begin{tabular}[c]{@{}c@{}}System\\ Call\\ Latency(s)\end{tabular}} 
     & getpid & 0.0065 & 0.0058 & 0.09\\
     & getrusage & 0.64 & 7.45 & 0.09 \\
     & gettimeofday & 0.0239 & 6.5986 & 6.85 \\
     & sbrk & 0.0064 & 0.0065 & 0.01 \\
     & sigaction & 2.21  & 2.79 & 0.59 \\
     & write & 0.51 & 7.23 & 0.53 \\ \midrule
    \multirow{2}{*}{\begin{tabular}[c]{@{}c@{}}Signal\\ Handler Latency\end{tabular}} 
    & Installing Signal & 2.24 & 2.79 & 0.60 \\
     & Handling Signal & 8.88 & 81.58 & 0.37 \\ 
    \bottomrule
    \end{tabular}
    }
    \caption{Summary of HBenchOS benchmark results for \gsgx along
    with \dr and \tool.}
    \label{tab:hbenchos-graphene}
    \end{table}  \subsection{Performance: Micro-benchmarks}
\label{sec:appx-bm}

We measure the performance for targeted workloads that stress system
APIs, CPU, and IO. The breakdown helps is to explain the costs
associated with executing diverse workloads with \codename.

\paragraph{Methodology}
For each target binary, we record the execution time in two following
settings:

\begin{itemize}
\item Baseline ({\dr}).
We execute the application binary directly with \dr.

\item {\codename}.
We use \codename to execute the application binary in enclave. We
offset the execution time by deducting the overhead to create,
initialize, load \dr and the application binary inside the enclave,
and destroy the enclave. 
\end{itemize}

\paragraph{System Stress Workloads.}
We use HBench-OS~\cite{hbenchos}---a benchmark to measure the
performance of primitive functionality provided by an OS and hardware
platform. In Table~\ref{tab:hbenchos} we show the cost of each
system-level operation such as system calls, memory operations,
context switches, and signal handling.
Memory-intensive operation latencies vary with benchmark setting: (a)
when the operations are done with more iterations (in millions) and
less memory chunk size ($4$ KB) the performance is comparable; (b)
when the operations are done with less iterations ($1$ K) and less
memory chunk size ($4$ MB) \codename incurs overhead ranging from
$117$\% to $651$\%. This happens because when the chunk size is large,
we need to allocate and de-allocate memory inside enclave for every
iteration as well as copy large amounts of data.

These file operation latencies match with latencies we observed in our
I/O intensive workloads (Figure~\ref{fig:iozone}). Specifically, the
write operation incurs large overhead. Hence, the $\tt{create}$
workload incurs $259$\% overhead because the benchmark creates a file
and then writes predefined sized data to it.
Costs of system calls that are executed as \ocalls vary depending on
return value and type of the system call. For example, system calls
such as $\tt{getpid, sbrk, sigaction}$ that return integer values
are much faster. Syscalls such as $\tt{getrusage, gettimeofday}$
returns structures or nested structures. Thus, copying these
structures back and forth to/from enclaves causes much of the
performance slowdown.
\codename has a custom mechanism for registering and handling signal
(Section~\ref{sec:sighand}); it introduces a latency of $19.71$\% and
$89.11$\% respectively. Registering signals is cheaper because it does
not cause a context switch as in the case of handling the signal. Further, after
accounting for the \ocall costs, our custom forwarding mechanism does
not introduce any significant slowdown.

\paragraph{CPU-bound Workloads.}
\codename incurs $40.24$\% overhead averaged over $23$ applications
from SPEC 2006~\cite{spec} with respect to \dr. Table~\ref{tab:stats}
shows the individual overheads for each application with respect to
all baselines. From Table~\ref{tab:stats} we observe that applications
that incur higher number of page faults and \ocalls suffer larger
performance slow-downs. Thus, similar to other SGX frameworks, the
costs of enclave context switches and limited EPC size are the main
bottlenecks in \codename.

\paragraph{IO-bound Workloads.}
\codename performs \ocalls for file I/O by copying read and write
buffers to and from enclave. We measure the per-API latencies using
FSCQ suite for file operations~\cite{fscq}.
Table~\ref{tab:stats} shows the costs of each
file operation and file access patterns respectively. Apart from the
cost of the \ocall, writes are more expensive compared to reads in
general; the multiple copy operations in \codename amplify the
performance gap between them.
Next we use IOZone~\cite{iozone}, a commonly used benchmark to measure
the file I/O latencies. Figure~\ref{fig:iozone} shows the bandwidth
over varied file sizes between $16$ MB to $1024$ MB and record sizes
between $4$ KB to $4096$ KB for common patterns. The trend of writes
being more expensive holds for IOZone too. \codename incurs an average
slowdown of $75$\% over all operations, record sizes, and file sizes.

\paragraph{Multi-threaded Workloads.}
We use the standard Parsec-SPLASH2~\cite{parsec-splash2} benchmark
suite. It comprises a variety of high performance computing (HPC) and
graphics applications.  We use it to benchmark \codename overheads for
multi-thread applications. Since some of the programs in
Parsec-SPLASH2 mandate the thread count to be power to $2$ (e.g.,
ocean\_ncp), we fixed the maximum number of threads in our experiment
to $16$. \codename changes the existing SGX design to handle thread
creation and synchronization primitives, as described in
Section~\ref{sec:threading} and ~\ref{sec:synch}. We measure the
effect of this specific change on the application execution by
configuring the enclave to use varying number of threads between
$1$-$16$.

Figure~\ref{fig:parsec} shows a performance overhead of
$832$\%, on average, across all benchmarks and thread configurations.
For single-threaded execution, \codename causes an overhead of $28$\%.
With the increase in threads, it varies from $416$\%, $910$\%,
$1371$\%, $1438$\% respectively.
We measure the breakdown of costs and observe that, on average: (a)
creating each thread contributes to a fixed cost of $57$ ms; 
(b) shared access to variables becomes expensive by a factor of $1-7$
times compared to the elapsed time of $\tt{futex}$ synchronization
with increase in number of threads.
This is expected because synchronization is cheaper in \dr execution,
in which it uses unsafe $\tt{futex}$ primitives exposed by the
kernel. On the other hand, \codename uses expensive spinlock mechanism
exposed by SGX hardware for security. Particularly, some of the
individual benchmarks, such as $\tt{water\_spatial}$, $\tt{fmm}$ and
$\tt{raytrace}$ that involve lots of lock contention events and have
extremely high frequency of spinlock calls (e.g., the spinning counts
of about $423,000$ ms in \codename while the $\tt{futex}$ calls of about
$500$ ms in \dr for the raytrace with $8$ threads). Thus, they incur 
large overheads in synchronization.

 \subsection{Performance:Real-world Case-studies}
\label{sec:appx-apps}

We work with $4$ representative real-world applications: a database
server (SQLite), a command-line utility (cURL), a machine learning
inference-as-a-service framework (Privado), and a key-value store
(memcached). These applications have been used in prior
work~\cite{panoply}.

\paragraph{SQLite} is a 
popular database~\cite{sqlite}. We select it as a case-study because
of its memory-intensive workload. We configure it as a single-threaded
instance. We use a database benchmark workload~\cite{leveldb} and
measured the throughput (ops/sec) for each database operation with
varying sizes of the database (total number of entries).
Table~\ref{tab:stats} shows the detailed breakdown of the runtime
statistics for a database with $10,000$ entries.
Figure~\ref{fig:sqlite} shows the average throughput over all
operations.
With \codename, we observe a throughput loss of $25.14$\% on average
over all database sizes.
The throughput loss increases with increase in the database size. The
drop is noticeable at $500 K$ where the database size crosses the
maximum enclave size threshhold and results in significant number of
page faults. This result matches with observations from other SGX
frameworks that report SQLite performance~\cite{scone}.

\paragraph{cURL}
is a widely used command line utility to download data from
URLs~\cite{curl}. It is network intensive. We test it with \codename
via the standard library test suite. Table~\ref{tab:stats} shows
detailed breakdown of time execution time on \codename. We measure the
cost executing cURL with \codename for downloading various sizes of
files from an Apache ($2.4.41$) server on the local network.
Figure~\ref{fig:curl} shows the throughput for various baselines and
file sizes. On average, \codename causes a loss of $142.27$\%
throughput as compared to \dr. For all baselines, small files (below
$100$ MB) have smaller download time; larger file sizes naturally take
longer time. This can be explained by the direct copying of packets
to non-enclave memory, which does not add any memory pressure on the
enclave. The only remaining bottleneck in the cost of dispatching
\ocalls which increase linearly with the requested file size.

\paragraph{Privado} 
is a machine learning framework that provides secure
inference-as-a-service~\cite{privado}. It comprises of several state
of the art models available as binaries that can execute on an input
image to predict its class. The binaries are CPU intensive and have
sizes ranging from $313$ KB to $140$ MB (see Table~\ref{tab:stats}).
We execute models from Privado on all the images from the
corresponding image dataset (CIFAR or ImageNet) and measure inference
time. 
Figure~\ref{fig:privado} shows the performance of baselines and \codename for $9$ models
in increasing order of binary size.
We observe that \codename performance degrades with increase in binary
size. This is expected because the limited enclave physical memory
leads to page faults. Hence, largest model ($140$ MB) exhibit highest
inference time and smallest model ($313$ KB) exhibit lowest inference
time. Thus, \codename and enclaves in general can add significant
overheads, even for CPU intensive server workloads, if they exceed the
working set size of $90$ MB.

\paragraph{Memcached}
is an in-memory key-value cache. We evaluate it with YCSB's all four
popular workloads A ($50\%$ read and $50\%$ update), B ($95\%$ read
and $5\%$ update), C ($100\%$ read) and D ($95\%$ read and $5\%$
insert). We run it with $4$ default worker threads running in \dr and
\codename settings. We vary the YCSB client threads with {\em Load}
and {\em Run} operations (to load the data and then run the
workload tests, respectively). We fix the data size to $1,000,000$
with Zipfian distribution of key popularity. We increase the number of
clients from $1$ to $100$ to find out a saturation point of each
targeted/scaled throughput for the settings. Here, we only present
workload A (throughput vs average latency for the read and update);
the other workloads display similar behavior.

As shown in Figure~\ref{fig:memcached}, the client latencies of the
two settings for a given throughput are slightly similar until
approximately $10,000$ ops/sec. Then. \codename jitters until it
achieves maximum throughput around $17,000$ ops/sec, while \dr is flat
until $15,000$ ops/sec (the maximum is $21,000$ operations per
second). The shared reason of the deceleration for both is that \dr 
slows down the speed of {\em Read} and {\em Update}. For
\codename, the additional bottleneck is the high frequency of lock
contention with spin-lock primitive. For e.g., \codename costs
$18,320,000$ ms while \dr's the futex calls cost only around $500$ ms
for a given throughput of $10000$ with $10$ clients.

\end{document}